\documentclass{PoS}

\newcommand{\ls}{{\stackrel{\textstyle <}{_\sim}}}
\newcommand{\msun}{{{M}_{\odot}}}

\newcommand{\fmmt}{{\rm fm}^{-3}}
\newcommand{\mevt}{{\rm MeV/fm}^3}

\newcommand{\okgr}{{\Omega_{\rm K}}}
\newcommand{\nuk}{{\nu_{\rm K}}}
\newcommand{\bag}{{B^{1/4}}}
\newcommand{\gcmt}{{\rm g/cm}^3}
\newcommand{\pkgr}{{P_{\,\rm K}}}

\newcommand{\kFn}{{k_{F_n}}}
\newcommand{\fmmo}{{\rm fm}^{-1}}

\newcommand{\gthree}{${\rm G}_{300}^{\rm B180 }$}
\newcommand{\hv}{HV}
\newcommand{\lsim}{\stackrel{\textstyle <}{_\sim}}

\newcommand{\icrust}{I_{\rm crust}}
\newcommand{\itotal}{I_{\rm total}}
\newcommand{\kFchi}{k_{F_\chi}}
\newcommand{\ecrusti}{\epsilon_{\rm crust}}
\newcommand{\fm}{{\rm fm}}
\newcommand{\const}{{\rm const}}
\newcommand{\KBt}{{\rm G}_{\rm 300}^{\rm B180}}
\newcommand{\kFp}{k_{F_p}}
\newcommand{\kFe}{k_{F_e}}
\newcommand{\bn}{$\begin{array}{l} \vspace{-0.2cm}}
\newcommand{\en}{\end{array}$}

\title{Strangeness in compact stars }

\ShortTitle{Strangeness in compact stars }

\author{\speaker{Fridolin Weber}\thanks{Supported by the National
Science Foundation under Grant PHY-0457329, and by the Research
Corporation.}\\ Department of Physics, San Diego State University \\
E-mail: \email{fweber@sciences.sdsu.edu}}

\author{Andreu Torres i Cuadrat\\
        Physics Department, Universitat Autonoma de Barcelona \\
        E-mail: \email{andreu.torres@uab.es}}

\author{Alexander Ho\\
        Department of Physics, San Diego State University \\
        E-mail: \email{aho@rohan.sdsu.edu}}

\author{Philip Rosenfield\\ Department of Astronomy, San Diego State
        University \\ E-mail: \email{philrose@sciences.sdsu.edu}}

\abstract{Astrophysicists distinguish between three different types of
compact stars. These are white dwarfs, neutron stars, and black
holes. The former contain matter in one of the densest forms found in
the Universe. This feature, together with the unprecedented progress
in observational astronomy, makes such stars superb astrophysical
laboratories for a broad range of exciting physical studies. This
article studies the role of strangeness for compact star
phenomenology. Strangeness is carried by hyperons, mesons,
H-dibaryons, and strange quark matter, and may leave its mark in the
masses, radii, cooling behavior, surface composition and the spin
evolution of compact stars.}

\FullConference{29th Johns Hopkins Workshop on current problems in
                 particle theory: strong matter in the heavens\\ 1-3
                 August\\ Budapest}

\begin{document}

\section{Introduction}

Astrophysicists distinguish between three different types of compact
stellar objects. These are white dwarfs, neutron stars, and black
holes. The latter constitute a region of space which has so much
mass--energy concentrated in it that no particles (not even light)
inside the black hole's event horizon can escape the black hole's
gravitational pull.  The situation is very different for neutron stars
and white dwarfs, which are about as massive as the sun (mass
$\msun=2\times 10^{30}$~kg) but whose radii are much smaller than the
sun's radius ($R_\odot = 7\times 10^5$~km).  Model calculations
predict that the matter in the cores of neutron stars is compressed to
densities ranging from a few times the density of an atomic nucleus,
$2.5\times 10^{14}$ g/cm$^3$, to densities that may be ten to twenty
times higher \cite{glen97:book,weber99:book,lattimer01:a}. In
comparison to that, white dwarfs of average mass, $M \sim 0.6 \,
\msun$ are at least by a factor of $10^7$ less dense than neutron
stars. The tremendous densities (and thus pressures) existing inside
neutron stars make them superb astrophysical laboratories for a wide
range of fascinating physical studies
\cite{glen97:book,weber99:book,weber05:a}. These include the
exploration of nuclear processes in an environment extremely rich of
electrons and neutrons, and the formation of new states of matter,
like quark matter which is being sought at the most powerful
terrestrial particle colliders. If quark matter exists in the cores of
neutron stars, it will be a color superconductor whose condensation
pattern has been shown to be very complex
\cite{rajagopal01:a,alford01:a,rischke05:a}. It has also been
theorized that quark matter (known as strange quark matter) may be
even more stable than atomic nuclei. In the latter event neutron stars
should be entirely made of strange quark matter, possibly enveloped in
a very thin nuclear crust. Such objects are called strange stars
\cite{alcock86:a,alcock88:a,madsen98:b}.  Strangeness, therefore,
carried by hyperons, mesons, H-dibaryons, and strange quark matter,
plays a key role for compact star physics and phenomenology, as will
be discussed in this paper (see Ref.\ \cite{weber05:a} for a detailed
recent review on this topic).

\section{Composition of high-density neutron star matter and EoS}
\label{sec:scomp}

\begin{figure}[htb]
\begin{center}
\includegraphics*[width=0.6\textwidth,angle=0,clip]{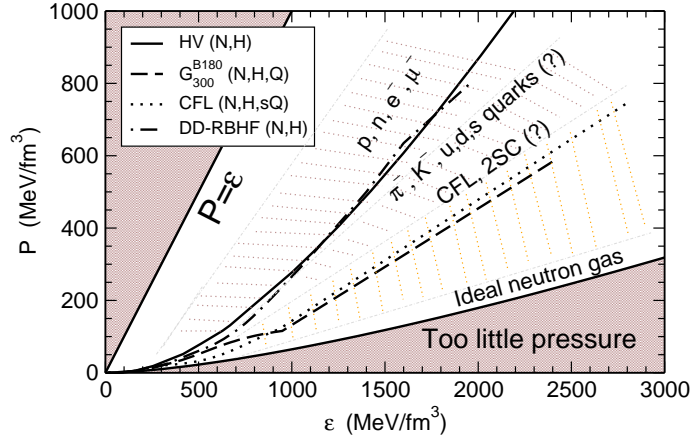}
\caption{Models for the EoS of neutron star matter.}
\label{fig:eos}
\end{center}
\end{figure}

The properties of neutron stars are determined by the equation of
state (EoS) of neutron star matter.  The EoS of neutron star matter
below neutron drip, which occurs at densities around $4\times
10^{11}\,\gcmt$, and at densities above neutron drip but below the
saturation density of nuclear matter is relatively well known.  This
is to a less extent the case for the EoS in the vicinity of the
saturation density of normal nuclear matter, $n_0=0.16~\fmmt$
(energy density of $\epsilon_0 = 140~\mevt$). Finally, the physical
properties of matter at still higher densities are extremely uncertain
so that the associated EoS is only very poorly known
\cite{glen97:book,weber99:book,lattimer01:a}. This is graphically
indicated by the hatched areas in Fig.\ \ref{fig:eos}, which is based
on three Walecka-type models for the EoS. The HV model is a
relativistic non-linear mean-field equation of state computed for
nucleons (N$=n,~ p$) and hyperons (H$= \Sigma, ~\Lambda, ~\Xi$) in
chemical equilibrium \cite{weber05:a,glen85:b}.  The other two models,
$\KBt$ \cite{glen97:book} and CFL \cite{alford03:b}, account
additionally for the presence of up, down, and strange quarks in
neutron star matter. The quarks are treated as normal unpaired quarks
(Q) in the $\KBt$ model, and as color-flavor locked superconducting
quarks (sQ) with a superfluid gap of $\Delta = 100$~MeV in the CFL
model. The model labeled DD-RBHF is a density dependent relativistic
Brueckner-Hartree-Fock EoS which accounts for nucleons and hyperons
\cite{lenske95:a,hofmann01:a}. It is obvious from Fig.\ \ref{fig:eos}
that, depending on stellar composition, neutron star properties such
as masses, radii, moments of inertia, redshifts, or limiting
rotational periods may vary significantly with strangeness.

\subsection{Hyperons}

At the densities in the interior of neutron stars, the neutron
chemical potential, $\mu^n$, is likely to exceeds the masses, modified
by interactions, of $\Sigma,~ \Lambda$ and possibly $\Xi$ hyperons
\cite{glen85:b}. Hence, in addition to nucleons, neutron star matter
is expected to have significant populations of hyperons and possibly
even $\Delta$'s \cite{huber98:a}. If so, pure neutron matter would
constitute an excited state relative to hyperonic (many-baryon) matter
which, therefore, would quickly transform via weak reactions like
\begin{equation}
n \rightarrow p + e^- + {\bar{\nu}}_e 
\label{eq:cnp}
\end{equation} to the lower energy state. The chemical 
potentials associated with reaction (\ref{eq:cnp}) in equilibrium obey the
relation
\begin{equation}
\mu^n = \mu^p + \mu^{e^-} \, ,
\label{eq:mun}
\end{equation} 
where $\mu^{\bar\nu_e}=0$ since the mean free path of (anti) neutrinos
is much smaller than the radius of neutron stars. Hence (anti)
neutrinos do not accumulate inside neutron stars. This is different
for hot proto-neutron stars \cite{prakash97:a}.
Equation~(\ref{eq:mun}) is a special case of the general relation
\begin{equation}
\mu^\chi = B^\chi \mu^n - q^\chi \mu^{e^-} \, , 
\label{eq:mui}
\end{equation} which holds in any system characterized by two conserved
charges. These are in the case of neutron star matter electric charge,
$q^\chi$, and baryon number charge, $B^\chi$. Application of Eq.\
(\ref{eq:mui}) to the $\Lambda$ hyperon ($B^\Lambda=1$, $q^\Lambda=0$), for
instance, leads to
\begin{equation}
\mu^\Lambda = \mu^n \, .
\label{eq:L}
\end{equation} Ignoring particle interactions, the chemical potential of a
relativistic particle of type $\chi$ is given by 
\begin{equation}
\mu^\chi = \omega(\kFchi) \equiv \sqrt{m_\chi^2 + \kFchi^2} \, ,
\label{eq:mukF}
\end{equation} where $\omega(\kFchi)$ is the single-particle 
energy of the particle and $\kFchi$ its Fermi momentum.  Substituting
(\ref{eq:mukF}) into (\ref{eq:L}) leads to
\begin{equation}
  \kFn \geq \sqrt{m_\Lambda^2 - m_n^2} \simeq 3~\fmmo \Rightarrow
    n \equiv { {\kFn^3}\over{3 \pi^2} } \simeq 2 n_0 \, ,
\label{eq:kFn}
\end{equation} 
where $m_\Lambda=1116$~MeV and $m_n=939$~MeV was used. That is, if
interactions among the particles are ignored, neutrons are replaced
with $\Lambda$'s in neutron star matter at densities as low as two
times the density of nuclear matter. This result is only slightly
altered by the inclusion of particle interactions \cite{glen85:b}.
Aside from chemical equilibrium, the condition of electric charge
neutrality of neutron star matter,
\begin{equation}
  \sum_{\chi=p, \Sigma^\pm, \Xi^-, \Delta^{++} , ...;
 e^-, \mu^-} q^\chi ~ {\kFchi^3} ~+ ~ 3 \, \pi^2 \, n^M \,
 \Theta(\mu^M - m_M) \equiv 0 \, ,
\label{eq:charge2}
\end{equation}
where $M$ stands for $\pi^-$ or $K^-$ mesons, plays a key role for the
particle composition of neutron star matter too. The last term in
(\ref{eq:charge2}) accounts for the possible existence of either a
$\pi^-$ or a $K^-$ meson condensate in neutron star matter, which will
be discussed in more detail in Sect.\ \ref{ssec:mcondens}
below. Before, however, we illustrate the importance of Eqs.\
(\ref{eq:mun}) and (\ref{eq:charge2}) for the proton-neutron fraction
of neutron star matter.  The beta decay and electron capture processes
among nucleons, $n \rightarrow p + e^- + \bar{\nu}_e$ and $p + e^-
\rightarrow n + \nu_e$ respectively, also known as nucleon direct Urca
processes, are only possible in neutron star matter if the proton
fraction exceeds a certain critical value
\cite{lattimer91:a}. Otherwise energy and momentum can not be
conserved simultaneously for these reactions so that they are
forbidden. For a neutron star made up of only nucleons and electrons,
it is rather straightforward to show that the critical proton fraction
is around $11\%$. This follows from ${\mathbf k}_{F_n} = {\mathbf
k}_{F_p} + {\mathbf k}_{F_e}$ combined with the condition of electric
charge neutrality of neutron star matter. The triangle inequality then
requires for the magnitudes of the particle Fermi momenta $k_{F_n}
\leq k_{F_p} + k_{F_e}$, and charge neutrality dictates that $\kFp =
\kFe$.  Substituting $\kFp = \kFe$ into the triangle inequality leads
to $\kFn \leq 2 \kFp$ so that for the particle number densities of
neutrons and protons $n_n \leq 8 n_p$.  Expressed as a fraction of the
system's total baryon number density, $n\equiv n_p + n_n$, one thus
arrives at $n_p / n > 1/9 \simeq 0.11$, which is the figure quoted
just above.  Medium effects and interactions among the particles
modify this value only slightly but the presence of muons raises it to
about $0.15$.  Hyperons, which may exist in neutron star matter rather
abundantly, produce neutrinos via direct Urca processes like $\Sigma^-
\rightarrow \Lambda + e^- + \bar{\nu}_e$ and $\Lambda + e^-
\rightarrow \Sigma^- + \nu_e$ \cite{prakash92:a}.  The direct Urca
processes are of key importance for neutron star cooling, which will
be discussed briefly in Sect.\ \ref{sec:cooling}. In most cases, the
nucleon direct Urca process is more efficient than the ones involving
hyperons \cite{haensel94:a,schaab95:a}.

\subsection{Meson condensates}\label{ssec:mcondens}

The pion or kaon meson fields may develop condensates in dense neutron
star matter. These condensates would have two important effects on
neutron stars. Firstly, they would soften the EoS above the critical
density for onset of condensation, which reduces the maximum neutron
star mass. Secondly, since the $<\!\!\pi^-\!\!>$ or $<\!\!K^-\!\!>$
condensates can absorb as little or as much momentum as required by
the scattering processes $n + <\!\!\pi^-\!\!> \rightarrow n + e^- +
\bar\nu_e$ or $n + <\!\!K^-\!\!> \rightarrow n + e^- + \bar\nu_e$, the
associated neutrino emissivities are very high which leads to fast
neutron star cooling \cite{schaab95:a,page05:a} (see Sect.\
\ref{sec:cooling}).  Since the $K^-$ condensate process involves a
change in strangeness, it is roughly by a factor $\sin^2 \theta_C
\simeq 1/20$ ($\theta_C$ denotes the Cabibbo angle) less efficient
than the $\pi^-$ condensate process.  However, medium effects can
reduce the impact of the $\pi^-$ condensate on stellar cooling by
about one order of magnitude, making it comparable to the efficiency
of the $K^-$ condensate.  Estimates predict the onset of $\pi^-$
condensation at densities around $n^\pi \sim 2 n_0$, with $n_0 = 0.16~{\rm
fm}^{-3}$ the empirical nuclear matter density.  However, this density
is very sensitive to the strength of the effective nucleon
particle-hole repulsion in the isospin $T=1$, spin $S=1$ channel,
which tends to suppress $\pi^-$ condensation and may push $n^\pi$ to
much higher values.  Similarly, depending on the nuclear model, the
threshold density for the onset of $K^-$ condensation, $n^K$, is
probably at least as high as $4 n_0$ \cite{waas97:a,glen99:a}.

\begin{figure}[htb]
\begin{center}
\includegraphics*[width=0.6\textwidth,angle=0]{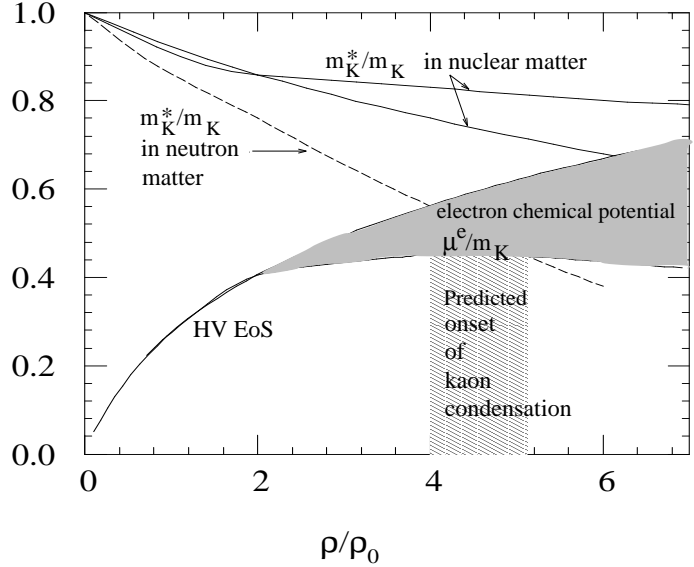}
\caption{The effective kaon mass, $m_K^*$, in nuclear matter and
  neutron star matter \cite{weber05:a}. Data taken from \cite{mao99:a}
  and \cite{waas97:a}, respectively.}
\label{fig:Kmass}
\end{center}
\end{figure}

$K^-$ condensation can only occur in neutron star matter if the
electron chemical potential equals the effective in-medium meson mass,
according to the schematic reaction $e^- \rightarrow K^- + \nu_e$,
with the neutrinos leaving the star (see Fig.\ \ref{fig:Kmass}). This
reaction would be followed
by $n \rightarrow p + K^-$. By this conversion the nucleons in the
cores of newly formed neutron stars can become half neutrons and half
protons \cite{brown96:a}. The relatively isospin symmetric composition
achieved in this way resembles the one of isospin symmetric atomic
nuclei, which are made up of equal numbers of neutrons and
protons. Neutron star matter is therefore referred to in this picture
as nucleon matter, and neutron stars constructed for such an EoS are
referred to as nucleon stars \cite{brown96:a,li97:a,li97:b,brown97:a}.

\subsection{H-dibaryons}

\begin{figure}[tb]
\begin{center}
\includegraphics[height=0.3\textheight]{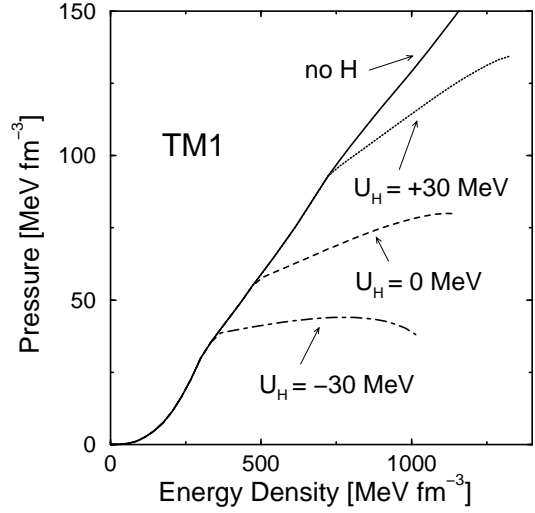}
\caption{EoS of neutron star matter accounting for a H-dibaryon
  condensate \cite{glen98:a}. $U_{\rm H}$ is the optical potential of
  the H-dibaryon at normal nuclear density.}
\label{fig:eosH}
\end{center}
\end{figure}

A novel particle that could make its appearance in the center of a
neutron star is the so-called H-dibaryon, a doubly strange six-quark
composite with spin and isospin zero, and baryon number two
\cite{jaffe77:a}. Since its first prediction in the 1970s, the
H-dibaryon has been the subject of many theoretical and experimental
studies as a possible candidate for a strongly bound exotic state.  In
neutron stars, which may contain a significant fraction of $\Lambda$
hyperons, the $\Lambda$'s could combine to form H-dibaryons, which
could give way to the formation of H-dibaryon matter at densities
somewhere above $\sim 3\, \epsilon_0$
\cite{tamagaki91:a,sakai97:a,glen98:a} depending on the in-medium
properties of the H-dibaryon.  For an attractive optical potential,
$U_{\rm H}$, of the H-dibaryon at normal nuclear density the equation
of state is softened considerably, as shown in Fig.\ \ref{fig:eosH}.
H-dibaryon matter could thus exist in the cores of moderately dense
neutron stars.  H-dibaryons with a vacuum mass of about 2.2~GeV and a
moderately attractive potential in the medium of
about $U_{\rm H} = - 30$~MeV, for instance, could go into a boson
condensate in the cores of neutron stars if the limiting star mass is
about that of the Hulse-Taylor pulsar PSR~1913+16, $M=1.444\, \msun$
\cite{glen98:a}. Conversely, if the medium potential were moderately
repulsive, around $U_{\rm H} = + 30$~MeV, the formation of H-dibaryons
may only take place in heavier neutron stars of masses greater than
about $1.6\, \msun$. If formed, however, H-matter may not remain
dormant in neutron stars but, because of its instability against
compression, could trigger the conversion of neutron stars into
hypothetical strange stars \cite{sakai97:a,faessler97:a,faessler97:b}.

\subsection{Quark deconfinement}\label{ssec:deconf}

One item that came recently into particular focus concerns the
possible existence of quark matter in the cores of neutron stars
\cite{weber99:book,rajagopal01:a,alford01:a}. The phase transition
from confined hadronic matter to deconfined quark matter is
characterized by the conservation of baryon charge and electric
charge. The Gibbs condition for phase equilibrium then is that the two
associated chemical potentials, $\mu^n$ and $\mu^e$, and the pressure
in the two phases be equal \cite{glen97:book,glen91:pt},
\begin{eqnarray}
  P_{\rm H}(\mu^n,\mu^e, \{ \chi \}, T) = P_{\rm Q}(\mu^n,\mu^e,T) \,
  .
\label{eq:gibbs1}
\end{eqnarray} $P_{\rm H}$ denoted the pressure of hadronic
matter computed for a given hadronic matter Lagrangian ${\cal L}_{\rm
M}(\{\chi\})$, where $\{\chi\}$ denotes the field variables and Fermi
momenta that characterize a solution to the field equations of
confined hadronic matter,
\begin{eqnarray}
  ( i \gamma^\mu\partial_\mu - m_\chi ) \psi_\chi(x) &=&
  \sum_{M=\sigma,\omega,\pi, ...} \Gamma_{M \chi} M(x) \, \psi_\chi(x) \, ,
   \\ 
 ( \partial^\mu\partial_\mu + m^2_\sigma) \sigma(x) &=&
  \sum_{\chi = p, n, \Sigma, ...} \Gamma_{\sigma \chi}\, \bar\psi_\chi(x)
  \psi_\chi(x) \, , 
\end{eqnarray}
 plus additional equations for the other meson fields ($M= \omega,
\pi, \rho , ...$).  The pressure of quark matter, $P_{\rm Q}$, is
obtainable from the bag model. The quark chemical potentials $\mu^u,
~\mu^d, ~\mu^s$ are related to the baryon and charge chemical
potentials as
\begin{eqnarray}
  \mu^u = {{1}\over{3}} \, \mu^n - {{2}\over{3}} \, \mu^e\,
  ,\qquad \mu^d = \mu^s = {{1}\over{3}} \, \mu^n + {{1}\over{3}} \,
  \mu^e \, .
\label{eq:cp.ChT}
\end{eqnarray} Equation~(\ref{eq:gibbs1}) is to be supplemented with
the two global relations for conservation of baryon charge and
electric charge within an unknown volume $V$ containing $A$
baryons. The first one is given by
\begin{equation}
  n \equiv {A\over V} = (1-\eta) \, n_{\rm H}(\mu^n,\mu^e,T) +
  \eta \, n_{\rm Q}(\mu^n,\mu^e,T) \, ,
\label{eq:bcharge}
\end{equation} where $\eta \equiv V_{\rm Q}/V$ denotes the volume
proportion of quark matter, $V_{\rm Q}$, in the unknown volume $V$, and
$n_{\rm H}$ and $n_{\rm Q} $ are the baryon number densities of hadronic
matter and quark matter.  Global neutrality of electric charge within the
volume $V$ can be written as
\begin{equation}
  0 = {Q\over V} = (1-\eta) \, q_{\rm H}(\mu^n,\mu^e,T) + \eta \,
  q_{\rm Q}(\mu^n,\mu^e,T)  +  q_{\rm L} \, ,
\label{eq:echarge}
\end{equation} with $q_i$ the electric charge densities of
hadrons, quarks, and leptons.  For a given temperature, $T$, Eqs.\
(\ref{eq:gibbs1}) to (\ref{eq:echarge}) serve to determine the two
independent chemical potentials and the volume $V$ for a specified
\begin{figure}[tb]
\begin{center}
\includegraphics*[width=0.9\textwidth,angle=0,clip]{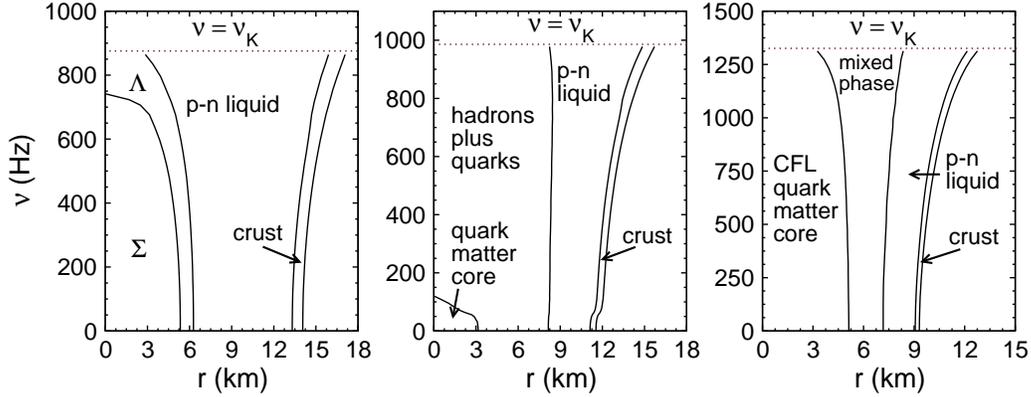}
\caption{Dependence of stellar compositions on neutron star spin
		 frequency, $\nu$, for the \hv, \gthree, and CFL (from
		 left to right) EoSs. The non-rotating stellar mass in
		 each case is about $1.4\, \msun$. $\nuk$ denotes the
		 Kepler (mass-shedding) frequency of each sequence,
		 discussed in Sect.\ \protect \ref{sssec:grav}.}
\label{fig:profiles}
\end{center}
\end{figure}
volume fraction $\eta$ of the quark phase in equilibrium with the
hadronic phase.  After completion $V_{\rm Q}$ is obtained as $V_{\rm
Q}=\eta V$. Because of Eqs.\ (\ref{eq:gibbs1}) through
(\ref{eq:echarge}) the chemical potentials depend on the proportion
$\eta$ of the phases in equilibrium, and hence so also all properties
that depend on them, i.e.\ the energy densities, baryon and charge
densities of each phase, and the common pressure. For the mixed phase,
the volume proportion of quark matter varies from $0 \leq \eta \leq
1$ and the energy density is the linear combination of the two
phases \cite{glen97:book,glen91:pt},
\begin{equation}
  \epsilon = (1-\eta) \, \epsilon_{\rm H}(\mu^n,\mu^e, \{\chi\}, T) + \eta \,
  \epsilon_{\rm Q}(\mu^n,\mu^e,T) \, .
\label{eq:eps.chi}
\end{equation} Hypothetical neutron star compositions computed 
along the lines described above are shown in Fig.\ \ref{fig:profiles}.
Possible astrophysical signals originating from quark deconfinement
will be discussed in Sect.\ \ref{sec:aps}
\cite{glen97:book,weber99:book,weber99:topr,glen00:b}.

\subsection{Color superconductivity of quark matter}

There has been much recent progress in our understanding of quark
matter, culminating in the discovery that if quark matter exists it
ought to be in a color superconducting state
\cite{rajagopal01:a,alford01:a,alford98:a,rapp98+99:a}. This is made
possible by the strong interaction among the quarks which is very
attractive in some channels. Pairs of quarks are thus expected to form
Cooper pairs very readily. Since pairs of quarks cannot be
color neutral, the resulting condensate will break the local color
symmetry and form what is called a color superconductor.  The phase
diagram of such matter is expected to be very complex
\cite{rajagopal01:a,alford01:a}.  This is caused by the fact that
quarks come in three different colors, different flavors, and
different masses.  Moreover, bulk matter is neutral with respect to
both electric and color charge, and is in chemical equilibrium under
the weak interaction processes that turn one quark flavor into
another. To illustrate the condensation pattern briefly, we note the
following pairing ansatz for the quark condensate \cite{alford03:a},
\begin{eqnarray}
\langle \psi^\alpha_{f_a} C\gamma_5 \psi^\beta_{f_b} \rangle \sim
\Delta_1 \epsilon^{\alpha\beta 1}\epsilon_{{f_a}{f_b}1} + \Delta_2
\epsilon^{\alpha\beta 2}\epsilon_{{f_a}{f_b}2} + \Delta_3
\epsilon^{\alpha\beta 3}\epsilon_{{f_a}{f_b}3} \, ,
\label{eq:pairing_ansatz}
\end{eqnarray}
where $\psi^\alpha_{f_a}$ is a quark of color $\alpha=(r,g,b)$ and
flavor ${f_a}=(u,d,s)$, and $\epsilon_{ijk}$ denotes the Levi-Civita
symbol. The latter is zero for $i=j$, $j=k$ , or $k=i$; $+1$ for
$(i,j,k)$ an even permutation of (1,2,3); and $-1$ for $(i,j,k)$ an
odd permutation of (1,2,3).  The condensate is a Lorentz scalar,
antisymmetric in Dirac indices, antisymmetric in color, and thus
antisymmetric in flavor. The gap parameters $\Delta_1$, $\Delta_2$ and
$\Delta_3$ describe $d$-$s$, $u$-$s$ and $u$-$d$ quark Cooper pairs,
respectively. The following pairing schemes have emerged. At
asymptotic densities ($m_s \rightarrow 0$ or $\mu \rightarrow \infty$)
the ground state of QCD with a vanishing strange quark mass is the
color-flavor locked (CFL) phase (color-flavor locked quark pairing),
in which all three quark flavors participate symmetrically.  The gaps
associated with this phase are
\begin{equation}
\Delta_3 \simeq \Delta_2 = \Delta_1 = \Delta \, ,
\label{eq:delta_CFL}
\end{equation}
and the quark condensates of the CFL phase are approximately of the form
\begin{equation}
 \langle \psi^{\alpha}_{f_a} C\gamma_5 \psi^{\beta}_{f_b} \rangle
\sim \Delta \, \epsilon^{\alpha \beta X} \epsilon_{{f_a} {f_b} X}
\, ,
\label{eq:CFL1}
\end{equation}
with color and flavor indices all running from 1 to 3. Since
$\epsilon^{\alpha\beta X} \epsilon_{{f_a} {f_b} X} =
\delta^\alpha_{f_a}\delta^\beta_{f_b} -
\delta^\alpha_{f_b}\delta^\beta_{f_a}$ one sees that the condensate
(\ref{eq:CFL1}) involves Kronecker delta functions that link color and
flavor indices. Hence the notion color-flavor locking. The CFL phase
has been shown to be electrically neutral without any need for
electrons for a significant range of chemical potentials and strange
quark masses \cite{rajagopal01:b}. If the strange quark mass is heavy
enough to be ignored, then up and down quarks may pair in the
two-flavor superconducting (2SC) phase.  Other possible condensation
patterns are CFL-$K^0$ \cite{bedaque01:a}, CFL-$K^+$ and
CFL-$\pi^{0,-}$ \cite{kaplan02:a}, gCFL (gapless CFL phase)
 \cite{alford03:a}, 1SC (single-flavor-pairing)
 \cite{alford03:a,buballa02:a,schmitt04:b}, CSL (color-spin locked
phase)  \cite{schaefer00:a}, and the LOFF (crystalline pairing)
 \cite{alford00:a,bowers02:a,casalbuoni04:a} phase, depending on $m_s$,
$\mu$, and electric charge density.  Calculations performed for
massless up and down quarks and a very heavy strange quark mass ($m_s
\rightarrow \infty$) agree that the quarks prefer to pair in the
two-flavor superconducting (2SC) phase where
\begin{equation}
\Delta_3 > 0\, , \quad {\rm and} \quad  \Delta_2 = \Delta_1 = 0 \, .
\label{eq:delta_2SC}
\end{equation}
In this case the pairing ansatz (\ref{eq:pairing_ansatz}) reduces to
\begin{equation}
 \langle \psi^{\alpha}_{f_a} C \gamma_5 \psi^{\beta}_{f_b} \rangle
\propto \Delta \, \epsilon_{ab} \epsilon^{\alpha \beta 3} \, .
\label{eq:2SC}
\end{equation}
Here the resulting condensate picks a color direction (3 or blue in
the example (\ref{eq:2SC}) above), and creates a gap $\Delta$ at the
Fermi surfaces of quarks with the other two out of three colors (red
and green). The gapless CFL phase (gCFL) may prevail over the CFL and
2SC phases at intermediate values of $m^2_s/\mu$ with gaps given
obeying the relation
\begin{equation}
\Delta_3 > \Delta_2 > \Delta_1 > 0 \, .
\label{eq:gCFL}
\end{equation}
For chemical potentials that are of astrophysical interest, $\mu <
1000$~MeV, the gap is between 50 and 100~MeV. The order of magnitude
of this result agrees with calculations based on phenomenological
effective interactions \cite{rapp98+99:a,alford99:b} as well as with
perturbative calculations for $\mu > 10$~GeV \cite{son99:a}. We also
note that superconductivity modifies the equation of state at the
order of $(\Delta / \mu)^2$ \cite{alford03:b,alford04:a}, which is
even for such large gaps only a few percent of the bulk energy. Such
small effects may be safely neglected in present determinations of
models for the equation of state of quark-hybrid stars. There has been
much recent work on how color superconductivity in neutron stars could
affect their properties (see  Refs.\
\cite{rajagopal01:a,alford01:a,alford00:a,rajagopal00:a,alford00:b,%
blaschke99:a} and references therein).  These studies reveal that
possible signatures include the cooling by neutrino emission, the
pattern of the arrival times of supernova neutrinos, the evolution of
neutron star magnetic fields, rotational stellar instabilities, and
glitches in rotation frequencies.

Aside from neutron star properties, an additional test of color
superconductivity may be provided by upcoming cosmic ray space
experiments such as AMS~\cite{ams01:homepage} and ECCO
\cite{ecco01:homepage}.\footnote{See J.\ Madsen's contribution on
strange matter in cosmic rays published elsewhere in this volume.} As
shown in Ref.~\cite{madsen01:a}, finite lumps of color-flavor locked
strange quark matter (see Sect.~\ref{ssec:ss}), which should be
present in cosmic rays if strange matter is the ground state of the
strong interaction, turn out to be significantly more stable than
strangelets without color-flavor locking for wide ranges of
parameters. In addition, strangelets made of CFL strange matter obey a
charge-mass relation of $Z/A \propto A^{-1/3}$, which differs
significantly from the charge-mass relation of strangelets made of
ordinary strange quark matter. In the latter case, $Z/A$ would be
constant for small baryon numbers $A$ and $Z/A \propto A^{-2/3}$ for
large $A$ \cite{madsen98:b,madsen01:a,aarhus91:proc}. This difference
may allow an experimental test of CFL locking in strange quark matter
\cite{madsen01:a}.

\subsection{Absolute stability of strange quark matter}
\label{ssec:ss}

So far we have assumed that quark matter forms a state of matter
higher in energy than atomic nuclei. This most plausible assumption,
however, may not be correct \cite{bodmer71:a,witten84:a,terazawa89:a},
since for a collection of more than a few hundred $u,~ d,~ s$ quarks,
the energy per baryon, $E/A$, of quark matter can be just as well
below the energy per baryon of the most stable atomic nuclei, nickel
and iron. This is known as the strange quark matter hypothesis. The
energy per baryon in $^{56}\rm{Fe}$, for instance, is given by
$M(^{56}{\rm Fe})c^2/56=930.4$~MeV, with $M(^{56}{\rm Fe})$ the mass
of the $^{56}$Fe atom.  A simple estimate shows that for strange
quark matter described by the MIT bag model $E/A = 4 B \pi^2/ \mu^3$,
so that bag constants of $B=57~\mevt$ (i.e.\ $\bag=145$~MeV) and
$B=85~\mevt$ ($\bag=160$~MeV) would place the energy per baryon of
strange quark matter at $E/A=829$~MeV and 915~MeV, respectively, which
correspond to strange quark matter which is absolutely bound with
respect to nuclear matter \cite{madsen98:b}. If this were indeed the
case, neutron star matter would be metastable with respect to strange
quark matter, and all neutron stars should in fact be strange quark
stars \cite{madsen98:b,madsen88:a,madsen93:a}. As outlined just above,
strange quark matter is expected to be a color superconductor which,
at extremely high densities, should be in the CFL phase.  This phase
is rigorously electrically neutral with no electrons required
\cite{rajagopal01:b}. For sufficiently large strange quark masses,
however, the low density regime of strange quark matter is rather
expected to form a 2SC phase (or possibly other phases) in which
electrons are present \cite{rajagopal01:a,alford01:a}. The presence of
electrons causes the formation of an electric dipole layer on the
surface of strange matter, which enables strange quark matter stars to
carry crusts made of ordinary nuclear matter
\cite{alcock86:a,alcock88:a,kettner94:b}. The maximal possible density
at the base of the crust (inner crust density) is determined by
neutron drip, which occurs at about $4\times 10^{11}~\gcmt$.  This
somewhat complicated situation of the structure of strange matter
enveloped in a (chemically equilibrated) nuclear crust can be
represented by a proper choice for the EoS which consists of two parts
\cite{glen92:crust}.
\begin{figure}
\begin{center}
\includegraphics[width=.6\textwidth,angle=0,clip]{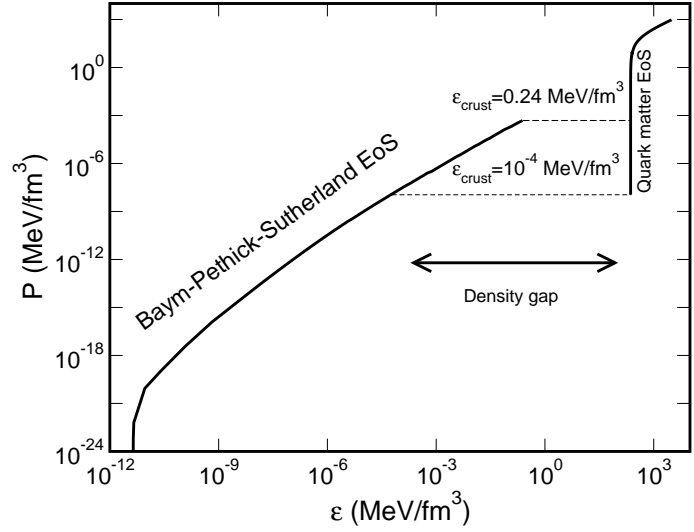}
\caption{EoS of strange quark matter surrounded by nuclear matter. The
maximal possible nuclear matter density is determined by neutron drip
which occurs at $\ecrusti = 0.24~\mevt$ ($4.3\times
10^{11}~\gcmt$). Any nuclear density that is smaller than that is
possible. As an example, we show here the EoS for a chosen density of
$\ecrusti = 10^{-4}~\mevt$ ($10^8~\gcmt$).}
\label{fig:eos.ss}
\end{center}
\end{figure}
At densities below neutron drip it can be represented by the EoS of
Baym-Pethick-Sutherland, while the high-density part, consisting of
strange quark matter, can be described by the bag model EoS (see Fig.\
\ref{fig:eos.ss}). The EoS is characterized by a significant
discontinuity in density between strange quark matter and nuclear
crust matter across the electric dipole gap where the pressure of the
nuclear crust at its base equals the pressure of strange matter at its
surface \cite{weber99:book,weber05:a,glen92:crust}.

One crucial astrophysical test of the strange quark matter hypothesis
is whether strange quark stars can give rise to the observed phenomena
of pulsar glitches. In the crust quake model an oblate solid nuclear
crust in its present shape slowly comes out of equilibrium with the
forces acting on it as the rotational period changes, and fractures
when the built up stress exceeds the sheer strength of the crust
material. The period and rate of change of period slowly heal to the
trend preceding the glitch as the coupling between crust and core
re-establish their co-rotation. The existence of glitches may have a
decisive impact on the question of whether the strange quark matter
hypothesis holds or not. From the calculations in \cite{glen92:crust}
it is known that the ratio of the crustal moment of inertia to the
star's total moment of inertia, $\icrust/\itotal$, varies between
$10^{-3}$ and $10^{-5}$ at the maximum mass.  If the angular
momentum of the pulsar is conserved in the quake then the relative
frequency change and moment of inertia change are equal and one
arrives at \cite{glen92:crust}
\begin{eqnarray}
       {{\Delta \Omega}\over{\Omega}} \; = \; 
       {{|\Delta I|}\over {I_0}} \; > \;
       {{|\Delta I|}\over {I}} \; \equiv \; f \;
       {\icrust\over I}\; \sim \; (10^{-5} - 10^{-3})\, f \; , 
       ~{\rm with}  \quad 0 < f < 1\; .
\label{eq:delomeg}
\end{eqnarray}
Here $I_0$ denotes the moment of inertia of that part of the star
whose frequency is changed in the quake. It might be that of the crust
only, some fraction, or all of the star. The factor $f$ in Eq.\
(\ref{eq:delomeg}) represents the fraction of the crustal moment of
inertia that is altered in the quake, i.e., $f \equiv |\Delta I|/
\icrust$.  Since the observed glitches have relative frequency changes
$\Delta \Omega/\Omega = (10^{-9} - 10^{-6})$, a change in the crustal
moment of inertia of $f\lsim 0.1$ would cause a giant glitch even in
the least favorable case \cite{glen92:crust}. Moreover, one finds that
the observed range of the fractional change in the spin-down rate,
$\dot \Omega$, is consistent with the crust having the small moment of
inertia calculated and the quake involving only a small fraction $f$
of that, just as in Eq.\ (\ref{eq:delomeg}).  To this aim we write
\cite{glen92:crust}
\begin{eqnarray}
        { {\Delta \dot\Omega}\over{\dot\Omega } } \; = \;
        { {\Delta \dot\Omega /  \dot\Omega} \over  
          {\Delta    \Omega  /     \Omega }  } \,
        { {|\Delta I |}\over{I_0} } \; = \; 
        { {\Delta \dot\Omega /  \dot\Omega} \over  
          {\Delta    \Omega  /     \Omega }  } \; f \;
         {\icrust\over {I_0} } \; > \; 
       (10^{-1}\; {\rm to} \; 10) \; f \; ,
\label{eq:omdot}
\end{eqnarray}
where use of Eq.\ (\ref{eq:delomeg}) has been made. Equation
(\ref{eq:omdot}) yields a small $f$ value in the range $f < (10^{-4}
\; {\rm to} \; 10^{-1})$, in agreement with $f\lsim 10^{-1}$
established just above. Here measured values of the ratio $(\Delta
\Omega/\Omega) / (\Delta\dot \Omega/\dot\Omega) \sim 10^{-6}$ to
$10^{-4}$ for the Crab and Vela pulsars, respectively, have been used.

An improved discussion of the surface gap below strange star crusts
has been performed very recently in Ref.\ \cite{stejner05:a}.  In
addition to the electrostatic forces described above, this study
includes gravity too. The properties of the gap are investigated for a
wide range of parameters assuming both color-flavor locked and
noncolor-flavor locked strange star cores.  It is found that the
maximally allowed inner crust density is generally lower than that of
neutron drip. This does not alter the overall form of the EoS shown in
Fig.\ \ref{fig:eos.ss}, however.  Another interesting finding
concerning the surface properties of strange stars was recently
published in Ref.\ \cite{jaikumar05:a}. In this paper it is found
that, depending on the surface tension of nuggets of strange matter, a
heterogeneous crust comprised of nuggets of strange quark matter
embedded in an uniform electron background may exist in the surface
region of strange stars. This heterogeneous strange star surface would
have a negligible electric field which would make the existence of an
ordinary nuclear crust, which requires a very strong electric field,
impossible.

\section{Models of compact stars}\label{sec:csm}

Neutron stars are objects of highly compressed matter so that the
geometry of space-time is changed considerably from flat space.  Thus
models of such stars are to be constructed in the framework of
Einstein's general theory of relativity combined with theories of
superdense matter. The effects of curved space-time are included by
coupling the energy-momentum density tensor for matter fields to
Einstein's field equations. The generally covariant Lagrangian
density is
\begin{equation}
{\cal L} = {\cal L}_{\rm E} + {\cal L}_{\rm G}  \, ,
\label{eq:L+G}
\end{equation}
where the dynamics of particles is introduced through the matter
Lagrangian ${\cal L}_{\rm M}$ added to the gravitational Lagrangian
${\cal L}_{\rm G}$. The latter is given by
\begin{equation}
{\cal L}_{\rm G} = g^{1/2} \,  R = g^{1/2} \, g^{\mu\nu} \, R_{\mu\nu} \, ,
\label{eq:L_G2}
\end{equation}
where $g^{\mu\nu}$ and $R_{\mu\nu}$ denote the metric tensor and the
Ricci tensor, respectively. The latter is given by
\begin{eqnarray}
  R_{\mu\nu} = \Gamma_{\mu\sigma, \, \nu}^\sigma - \Gamma_{\mu\nu, \,
  \sigma }^\sigma + \Gamma_{\kappa\nu}^\sigma \,
  \Gamma_{\mu\sigma}^\kappa - \Gamma_{\kappa\sigma}^\sigma \,
  \Gamma_{\mu\nu}^\kappa \, ,
\label{eq:14.27} 
\end{eqnarray} 
where commas followed by a Greek letter denote derivatives with respect
to space-time coordinates, e.g. ${,_\nu} = {{\partial}/{\partial
x^\nu}}$ etc.  The Christoffel symbols $\Gamma$ in (\ref{eq:14.27})
are defined as
\begin{eqnarray}
  \Gamma_{\mu\nu}^{\sigma} = {{1}\over{2}}\, g^{\sigma\lambda}\,
  \left( g_{\mu\lambda,\, \nu} + g_{\nu\lambda,\, \mu } - g_{\mu\nu,
  \, \lambda} \right)\, .
\label{eq:14.17}
\end{eqnarray} 
The connection between both branches of physics is provided by
Einstein's field equations
\begin{eqnarray}
  G^{\mu\nu} \equiv R^{\mu\nu} - {{1}\over{2}} g^{\mu\nu} R = 8 \pi
T^{\mu\nu}(\epsilon,P(\epsilon)) \, ,
\label{eq:intro.1}
\end{eqnarray}
($\mu, \nu= 0, 1, 2, 3$) which couples the Einstein curvature tensor,
$G^{\mu\nu}$, to the energy-momentum density tensor, $T^{\mu\nu}$, of
the stellar matter. The quantities $g^{\mu\nu}$ and $R$ in
(\ref{eq:intro.1}) denote the metric tensor and the Ricci scalar
(scalar curvature) \cite{weber99:book}.  The tensor $T^{\mu\nu}$
contains the equation of state, $P(\epsilon)$, of the stellar matter
discussed in Sect.\ \ref{sec:scomp}.  In general, Einstein's field
equations and the many-body equations were to be solved simultaneously
since the baryons and quarks move in curved space-time whose geometry,
determined by Einstein's field equations, is coupled to the total mass
energy density of the matter.  In the case of neutron stars, as for
all astrophysical situations for which the long-range gravitational
forces can be cleanly separated from the short-range forces, the
deviation from flat space-time over the length scale of the strong
interaction, $\sim 1~\fm$, is however practically zero up to the
highest densities reached in the cores of such stars (some
$10^{15}~\gcmt$).  This is not to be confused with the global length
scale of neutron stars, $\sim 10$~km, for which $M / R \sim 0.3$,
depending on the star's mass.  That is to say, gravity curves
space-time only on a macroscopic length scale but leaves it flat to a
very good approximation on a microscopic length scale. To achieve an
appreciable curvature on a microscopic scale set by the strong
interaction, mass densities greater than $\sim 10^{40}~\gcmt$ would be
necessary \cite{thorne66:a}!  This circumstance divides the
construction of models of compact stars into two distinct
problems. Firstly, the effects of the short-range nuclear forces on
the properties of matter are described in a comoving proper reference
frame (local inertial frame), where space-time is flat, by the
parameters and laws of
\begin{figure}[htb]
\begin{center}
\includegraphics*[width=0.8\textwidth,angle=0,clip]{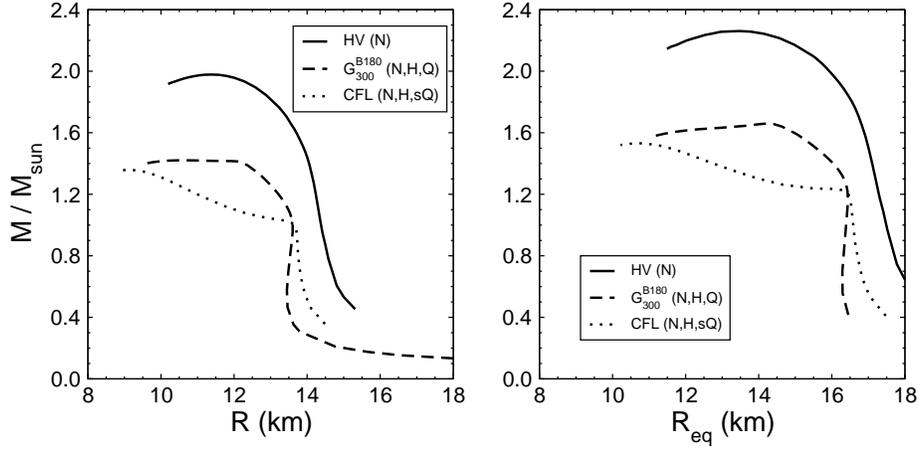}
\caption{Mass--radius relations of non-rotating (left panel) and
         rotating (right panel: $\Omega = \Omega_K$) neutron stars
         computed for different EoSs. $\Omega_K$ denotes the general
         relativistic Kepler (mass shedding) frequency.}
\label{fig:mvsr}
\end{center}
\end{figure}
special relativistic many-body physics.  Secondly, the coupling
between the long-range gravitational field and the matter is then
taken into account by solving Einstein's field equations for the
gravitational field described by the general relativistic curvature of
space-time, which determines the global structure of the stellar
configuration.

\begin{figure}[htb]
\begin{center}
\includegraphics*[width=0.8\textwidth,angle=0,clip]{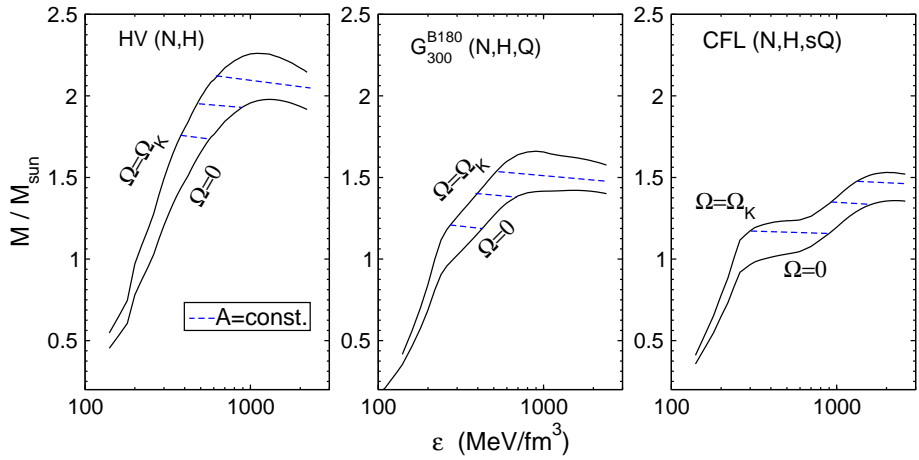}
\caption{Mass--central energy relations for the sample equations of
         state introduced in Sect.\ \protect \ref{sec:scomp}.}
\label{fig:mvse}
\end{center}
\end{figure}

For many studies of neutron star properties it is sufficient to treat
neutron star matter as a perfect fluid. The energy-momentum tensor of
such a fluid is given by
\begin{eqnarray}
  T^{\mu\nu} = {{dx^\mu}\over{d\tau}} \, {{dx^\nu}\over{d\tau}} \,
  \bigl(\, \epsilon + P \, \bigr) \, + \, g^{\mu\nu} \, P \, .
\label{eq:85.7}
\end{eqnarray}
For non-rotating spherically symmetric stars the metric has the rather
simple form
\begin{eqnarray}
  ds^2 = - e^{2\,\Phi(r)} \, dt^2 + e^{2\,\Lambda(r)} \, dr^2 +
  r^2 \, d\theta^2 + r^2 \, {\rm sin}^2\theta \, d\phi^2\, ,
\label{eq:15.20}
\end{eqnarray} where $\Phi(r)$ and $\Lambda(r)$ are radially
varying metric functions. From (\ref{eq:15.20}) one reads off the
following covariant components of the metric tensor,
\begin{equation}
g_{t t} = - \, {e^{2\,\Phi(r)}} \, , ~
g_{r r} = {e^{2\,\Lambda(r)}} \, , ~
g_{\theta \theta} =  {r}^{2} \, , ~
g_{\phi \phi} =  {r}^{2} \sin^2\!\theta \, ,
\label{eq:15.34}
\end{equation} 
so that the only non-vanishing Christoffel symbols are
\begin{eqnarray}
%    Chr(dn,dn,up)   [1, 1, 2]  
&&\Gamma_{t t}^r = 
 {e^{2\,\Phi(r)-2\,\Lambda(r)}} \, \Phi'(r) \, , ~
%    Chr(dn,dn,up)   [1, 2, 1]
\Gamma_{t r}^t =  \Phi'(r) \, , ~
%    Chr(dn,dn,up)   [2, 2, 2]
\Gamma_{r r}^r = \Lambda'(r)  \, , ~ 
%    Chr(dn,dn,up)   [2, 3, 3]
\Gamma_{r \theta}^\theta = {r}^{-1}  \, , ~
%    Chr(dn,dn,up)   [2, 4, 4]
\Gamma_{r \phi}^\phi =  {r}^{-1}  \, , ~
%    Chr(dn,dn,up)   [3, 3, 2] 
\Gamma_{\theta \theta}^r = - \, r \; e^{-2\,\Lambda(r)}  \, , 
\nonumber \\
%    Chr(dn,dn,up)   [3, 4, 4]
&& \Gamma_{\theta \phi}^\phi =  {\frac {\cos\theta}{\sin\theta}} \, , ~
%    Chr(dn,dn,up)   [4, 4, 2]
\Gamma_{\phi \phi}^r = - \, r \, \sin^2\!\theta \;
e^{-2\,\Lambda(r)}  \, , ~
%    Chr(dn,dn,up)   [4, 4, 3]
\Gamma_{\phi \phi}^\theta = -\sin\theta \, \cos\theta \, ,
\label{eq:15.54c}
\end{eqnarray} 
where primes denote differentiation with respect to the radial
coordinate. Substituting (\ref{eq:85.7}) and (\ref{eq:15.54c}) into
Einstein's field equations leads to the general relativistic equations
of hydrostatic equilibrium discussed first by Tolman \cite{tolman39:a}
and Oppenheimer-Volkoff \cite{oppenheimer39},
\begin{eqnarray}
{{dP(r)}\over{dr}} = - \, \frac{\epsilon(r)\, m(r)}{r^2} \;
\frac{\left( 1 + {{P(r)}\over{\epsilon(r)}} \right) \, 
\left( 1 + {{4 \pi r^3 P(r)}\over {m(r)}} \right)} 
{1 - {{2 m(r)}\over{r}}} \, .
\label{eq:f28}
\end{eqnarray}
Note that we use geometrized units, where the gravitational constant
and velocity of light are $G=c=1$ so that $\msun = 1.475$~km.  The
boundary condition to (\ref{eq:f28}) is $P(r=0) = P(\epsilon_c)$,
where $\epsilon_c$ denotes the energy density at the star's center,
which constitutes a parameter.  Equation (\ref{eq:f28}) is to be
integrated out to a radial distance where $P(r)=0$ which determines
the star's radius, $R$.  The mass contained in a sphere of radius
$r~(\leq R)$, denoted by $m(r)$, follows from $m(r) = 4 \pi \int^r_0
dr'\; r'^2 \; \epsilon(r') \, .$ The star's total gravitational mass
is thus given by $M\equiv m(R)$. Figure \ref{fig:mvsr} shows the
mass-radius relationships of both non-rotating as well as rotating
sequences of neutron stars for the sample EoSs discussed in Sect.\
\ref{sec:scomp}. The non-rotating sequences are solutions of the
Tolman--Oppenheimer--Volkoff equation shown in (\ref{eq:f28}). The
construction of rotating sequences will be discussed shortly below.
Figure \ref{fig:mvse} shows the gravitational mass of non-rotating as
well as rotating neutron stars as a function of central star density.
Stars to the right of the respective mass peaks in each panel are
unstable against radial oscillations and thus cannot exist stably in
nature. Also shown in these plots are the evolutionary (constant
stellar baryon number, $A$) paths that isolated rotating neutron stars
would follow during their stellar spin-down evolution caused by the
emission of magnetic dipole radiation and a wind of $e^+$--$e^-$
pairs. Figure \ref{fig:mvse} reveals that CFL stars may spend considerably
more time in the spin-down phase than their competitors of the same
mass.
Another point that we want to make is that all equations of state are
able to support neutron stars of canonical mass, $M \sim 1.4 \,
\msun$.  Neutron stars more massive than about $2\, \msun$, on the
other hand, are only supported by equations of state that exhibit a
very stiff behavior at asymptotic densities, disfavoring the presence
of hyperons, meson condensates, or quarks. Knowledge of the maximum
possible mass of neutron stars is of great importance for two
reasons. Firstly, because the largest known neutron star mass imposes
a lower bound on the maximum mass of a theoretical model. Very massive
neutron star candidates are J0751+1807 ($2.1^{+0.4}_{-0.5}\, \msun$
\cite{nice05:a}), Vela X--1 ($1.88\pm 0.13\, \msun$ if the inclination
angle of the system is $i=90^{\rm o}$; an inclination angle of
$i=70^{\rm o}$ increases the star's mass to $2.27\pm 0.17\, \msun$
\cite{quaintrell03:a}), and Cyg X-2 ($1.78\pm 0.23 \, \msun$
\cite{orosz99:a}. Titarchuck and Shaposhnikov obtain for Cyg X--2 a
lower mass of $1.44\pm 0.06 \, \msun$ \cite{titarchuck02:a}).  The
second reason is that the maximum mass of neutron stars is essential
in order to identify solar-mass black hole candidates
\cite{brown94:a,bethe95:a}.

The structure equations of rotating compact stars are considerably
more complicated that those of non-rotating compact stars
\cite{weber99:book}.  These complications have their cause in the
rotational deformation, that is, a flattening at the pole accompanied
with a radial blowup in the equatorial direction, which leads to a
dependence of the star's metric on the polar coordinate, $\theta$, in
addition to the mere dependence on the radial coordinate, $r$.
Secondly, rotation stabilizes a star against gravitational collapse. A
rotating star can therefore carry more mass than a non-rotating star.
Being more massive,
\begin{figure}[tb]
\begin{center}
\includegraphics*[width=0.6\textwidth,angle=0,clip]{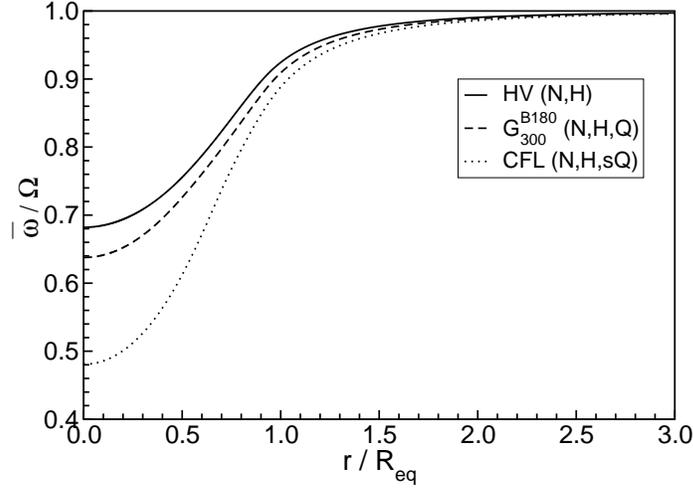}
\caption{Dragging of local inertial frames (Lense-Thirring effect)
  caused by $\sim 1.4 \, \msun$ neutron stars rotating at 2~ms.  The
  frequency $\bar\omega$ is defined in Eq.\ (\protect
  \ref{eq:bar.omega}).}
\label{fig:fd}
\end{center}
\end{figure}
however, means that the geometry of space-time is changed too. This
makes the metric functions associated with a rotating star depend on
the star's rotational frequency.  Finally, the general relativistic
effect of the dragging of local inertial frames implies the occurrence
of an additional non-diagonal term, $g^{t\phi}$, in the metric tensor
$g^{\mu\nu}$. This term imposes a self-consistency condition on the
stellar structure equations, since the extent to which the local
inertial frames are dragged along by the star is determined by the
initially unknown stellar properties like mass and limiting rotational
frequency. The covariant components of the metric tensor of a rotating
compact star are thus given by \cite{weber99:book,friedman86:a}
\begin{eqnarray}
g_{t t} = - {e^{2\,{\nu}}} + {e^{2\,\psi}} \omega^2 \, , ~ g_{t \phi}
 = - {e^{2\,\psi}} \omega \, , ~ g_{r r} = {e^{2\,\lambda}} \, , ~
 g_{\theta \theta} = {e^{2\,{\mu}}} \, , ~ g_{\phi \phi} =
 {e^{2\,\psi}} \, ,
\label{eq:11.4bk} 
\end{eqnarray}
which leads for the line element to
\begin{eqnarray}
  d s^2 = g_{\mu\nu} dx^\mu dx^\nu = - \, e^{2\,\nu} \, dt^2 +
e^{2\,\psi} \, \bigl( d\phi - \omega \, dt \bigr)^2 + e^{2\,\mu} \,
d\theta^2 + e^{2\,\lambda} \, dr^2 \, .
\label{eq:f220.exact} 
\end{eqnarray} Here each metric function, i.e.\ $\nu$, $\psi$,  $\mu$ and
$\lambda$, as well as the angular velocities of the local inertial
\begin{table}   %[H] add [H] placement to break table across pages
\caption{\label{tab:redshifts} Properties of neutron stars composed of
nucleons and hyperons (\hv), nucleons, hyperons, and normal quarks
(\gthree), and nucleons, hyperons, and color-superconducting quarks
(CFL).}
\begin{center}
\begin{tabular}{|c|c|c|c|c|c|c|} \hline

 & ~\hv~       & ~\gthree~  & ~CFL~     & ~\hv~       & ~\gthree~  & ~CFL~     \\
% & $\Omega$ = 0  &$\Omega = 0$  &$\Omega = 0$ & $\okgr$ = 5360~${\rm s^{-1}}$ 
% & $\okgr$ = 5900~${\rm s^{-1}}$ & $\okgr$ = 8800~${\rm s^{-1}}$ \\
& $\nu$ = 0  &$\nu = 0$  &$\nuk = 0$ & $\nuk$ = 850~Hz
& $\nuk$ = 940~Hz  &$\nuk$ = 1400~Hz \\
\hline \hline
$\epsilon_{\rm c}~(\mevt)$ & 361.0  & 814.3  &2300.0 &280.0  &400.0 &1100.0  \\
\hline
$I~{\rm (km^{3})}$         & 0      & 0      & 0      & 223.6   & 217.1   & 131.8   \\
\hline
$M$~($\msun$)                  & 1.39   & 1.40   & 1.36   & 1.39    & 1.40    & 1.41    \\
\hline
$R$~(km)                     & 14.1   & 12.2   & 9.0    & 17.1    & 16.0    & 12.6    \\
\hline
$Z_{\rm p}$           & 0.1889 & 0.2322 & 0.3356 & 0.2374  & 0.2646  & 0.3618  \\
$Z_{\rm F}$     & 0.1889 & 0.2322 & 0.3356 & $-$0.1788 & $-$0.1817 & $-$0.2184 \\
$Z_{\rm B}$             & 0.1889 & 0.2322 & 0.3356 & 0.6046  & 0.6502  & 0.9190  \\
\hline
$g_{\rm {s,14}~ (cm/s^{2})}$ & 1.1086 & 1.5447 & 3.0146 & 0.7278  & 0.8487  & 1.4493  \\
\hline
$T/W$                      & 0      & 0      & 0      & 0.0894  & 0.0941  & 0.0787  \\
\hline
$BE~(\msun)$               & 0.0937 & 0.1470 & 0.1534 & 0.0524  & 0.1097  & 0.1203  \\
\hline
$V_{\rm eq}/c$           & 0      & 0      & 0      & 0.336   & 0.353   & 0.424   \\
\hline
\end{tabular}
\end{center}
\end{table}
frames, $\omega$, depend on the radial coordinate $r$ and polar angle
$\theta$ and implicitly on the star's angular velocity $\Omega$.  Of
particular interest is the relative angular frame dragging frequency,
$\bar\omega$, defined as
\begin{equation}
  \bar\omega(r,\theta,\Omega) \equiv \Omega - \omega(r,\theta,\Omega)
  \, ,
\label{eq:bar.omega}
\end{equation} which is the angular velocity of the star, $\Omega$,
relative to the angular velocity of a local inertial frame,
$\omega$. It is this frequency that is of relevance when discussing
the rotational flow of the fluid inside the star, since the magnitude
of the centrifugal force acting on a fluid element is governed--in
general relativity as well as in Newtonian gravitational theory--by
the rate of rotation of the fluid element relative to a local inertial
frame \cite{hartle67:a}. In contrast to Newtonian theory, however, the
inertial frames inside (and outside) a general relativistic fluid are
not at rest with respect to the distant stars, as pointed out just
above.  Rather, the local inertial frames are dragged along by the
rotating fluid. Depending on the internal stellar constitution, this
effect can be quite strong, as shown in Fig.\ \ref{fig:fd} for
rotating 2~ms neutron stars. For a very compact neutron star
containing a color-superconducting CFL core, as in our example, one
reads off from this figure that the local inertial frames at the
star's center rotate at about half the star's rotational frequency,
$\omega(r=0) \simeq \Omega/2$. This value drops to about 15\% for the
local inertial frames located at the star's equator. The scenarios
shown in Fig.\ \ref{fig:fd} may be of great importance for binary
millisecond neutron stars in their final accretion stages, where the
accretion disk approaches the star very closely.

Table \ref{tab:redshifts} summarizes the impact of strangeness on
several intriguing properties of non-rotating as well as rotating
neutron stars. The latter spin at their respective Kepler frequencies.
One sees that the central energy density, $\epsilon_{\rm c}$, spans a
very wide range, depending on particle composition. The surface
redshift is of importance since it is connected to observed neutron
star temperatures through the relation $T^\infty/T_{\rm eff} = 1/(1+Z)$.
CFL quark stars may have redshifts that are up to 50\% higher than
those of conventional stars. Finally, we also show in Table
\ref{tab:redshifts} the surface gravity of stars, $g_{{\rm s},14}$
\cite{bejger04:a}, which again may be up to 50\% higher for CFL
stars. The other quantities listed are the rotational kinetic energy
in units of the total energy of the star, $T/W$,
the stellar binding energy, $BE$, and the rotational velocity of a
particle at the star's equator \cite{weber99:book}.

\section{Limiting rotational periods}\label{sssec:grav}

\subsection{Mass shedding from the equator}

No simple stability criteria are known for rapidly rotating stellar
configurations in general relativity. However, an absolute limit on
rapid rotation is set by the onset of mass shedding from the equator
of a rotating star. The corresponding rotational frequency is known as
the Kepler frequency, $\okgr$. In classical mechanics, the expression
for the Kepler frequency, determined by the equality between the
centrifugal force and gravity, is readily obtained as $\okgr =
\sqrt{M/R^3}$. In order to derive the general relativistic counterpart
of this relation, one applies the extremal principle to the circular
orbit of a point mass rotating at the star's equator.  Since
$r=\theta=\const$ for a point mass there, one has $d r=d\theta=0$.  The
line element (\ref{eq:f220.exact}) then reduces to
\begin{eqnarray}
  d s^2 = \left( e^{2\,\nu} - e^{2\,\psi} \, (\Omega -
\omega)^2 \right) \, d t^2 \, .
\label{eq:12.1bk}
\end{eqnarray} Substituting this expression into $J \equiv
\int^{s_2}_{s_1} d s$, where $s_1$ and $s_2$ refer to points located at
\begin{figure}[tb]
\begin{center}
\includegraphics*[width=0.6\textwidth,angle=0,clip]{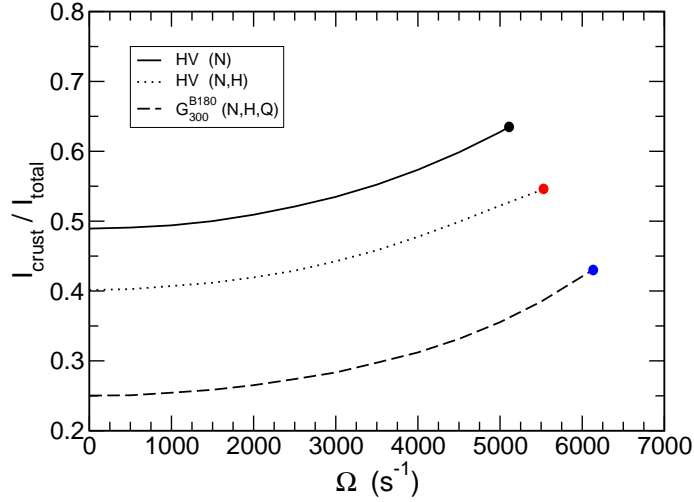}
\caption{Moment of inertia of several sample neutron stars.}
\label{fig:moi}
\end{center}
\end{figure}
that particular orbit for which $J$ becomes extremal, gives
\begin{eqnarray}
  J = \int_{s_1}^{s_2}\! d t \, \sqrt{e^{2\,\nu} - e^{2\,\psi}
    \, (\Omega - \omega)^2} \, .
\label{eq:12.2bk}
\end{eqnarray} Applying  the extremal condition $\delta J=0$ to
Eq.\ (\ref{eq:12.2bk}) and noticing that $V = e^{\psi-\nu} \, (\Omega
- \omega)$ then leads to the following relation,
\begin{eqnarray}
  {{\partial\psi}\over{\partial r}}\, e^{2\,\nu} \, V^2 -
  {{\partial\omega}\over{\partial r}}\, e^{\nu+\psi}\, V -
  {{\partial\nu}\over{\partial r}}\, e^{2\,\nu} = 0 \, . 
\label{eq:12.6bk}
\end{eqnarray} It constitutes a simple quadratic equation for the 
orbital velocity $V$ of a particle at the star's equator. One thus
obtains for the Kepler frequency $\okgr$ (Kepler period, $\pkgr$) the
final relation \cite{weber99:book},
\begin{eqnarray}
  \okgr = \omega +\frac{\omega^\prime}{2\psi^\prime} +e^{\nu -\psi} \sqrt{
    \frac{\nu^\prime}{\psi^\prime} + \Bigl(\frac{\omega^\prime}{2
      \psi^\prime}e^{\psi-\nu}\Bigr)^2 }  \quad 
\Rightarrow \quad \pkgr = {{2 \pi}
    \over {\okgr}} \, ,
\label{eq:okgr}  
\end{eqnarray} which is to be determined  self-consistently at the star's 
equator (primes denote radial derivatives). For most neutron star
matter equations of state, the Kepler period obtained for $1.4\,
\msun$ neutron stars scatters around 1~ms. One exception to this are
strange quark matter stars. These are self-bound and, thus, tend to
possess smaller radii than conventional neutron stars, which are bound
by gravity only. Because of their smaller radii, strange stars can
withstand mass shedding down to periods of around 0.5~ms
\cite{glen92:crust,glen92:limit}. CFL stars reside between these
limits.

As a last topic of this section, we briefly discuss the moment of
inertia of a rotationally deformed star described by the metric in
Eq.\ (\ref{eq:f220.exact}). For such stars the moment of inertia is
given by
\begin{eqnarray}
  I(\Omega) = 2\, \pi \int_0^\pi \! d\theta \int_0^{R(\theta)} d r \;
  e^{\lambda+\mu+\nu+\psi} \, {{\epsilon + P(\epsilon)}\over{e^{2\nu -
  2\psi} - (\omega - \Omega)^2}} \, {{\Omega - \omega}\over{\Omega}}
  \, . 
\label{eq:11.71bk} 
\end{eqnarray}
Figure~\ref{fig:moi} shows that the crustal fraction of the moment of
inertia of a neutron star may be around 50\% smaller
if the star contains a very soft phase of matter like quark
matter. This may be of relevance for pulsar glitch models and the
modeling of the post-glitch behavior of pulsars.

\subsection{Gravitational radiation reaction driven instabilities}

Rotational instabilities in rotating stars, known as gravitational
radiation driven instabilities, are probably setting a more stringent
limit on rapid stellar rotation than mass shedding.  These
instabilities originate from counter-rotating surface vibrational
modes which at sufficiently high rotational star frequencies are
dragged forward. In this case gravitational radiation, which
inevitably accompanies the aspherical
\begin{figure}[tb]
\begin{center} 
\includegraphics[width=7.0cm,angle=0]{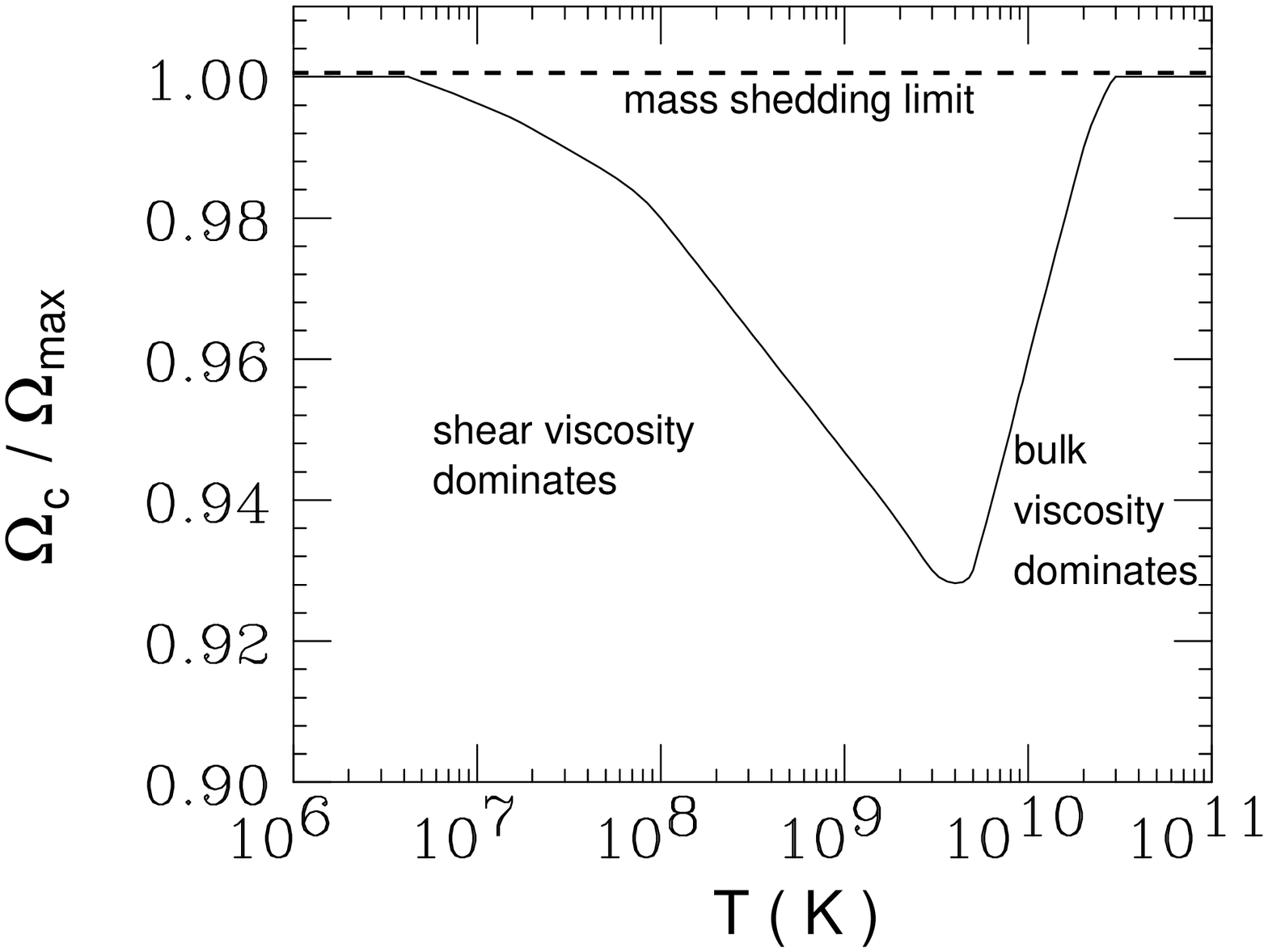} \hskip 1.0cm
\includegraphics[width=7.0cm,angle=0]{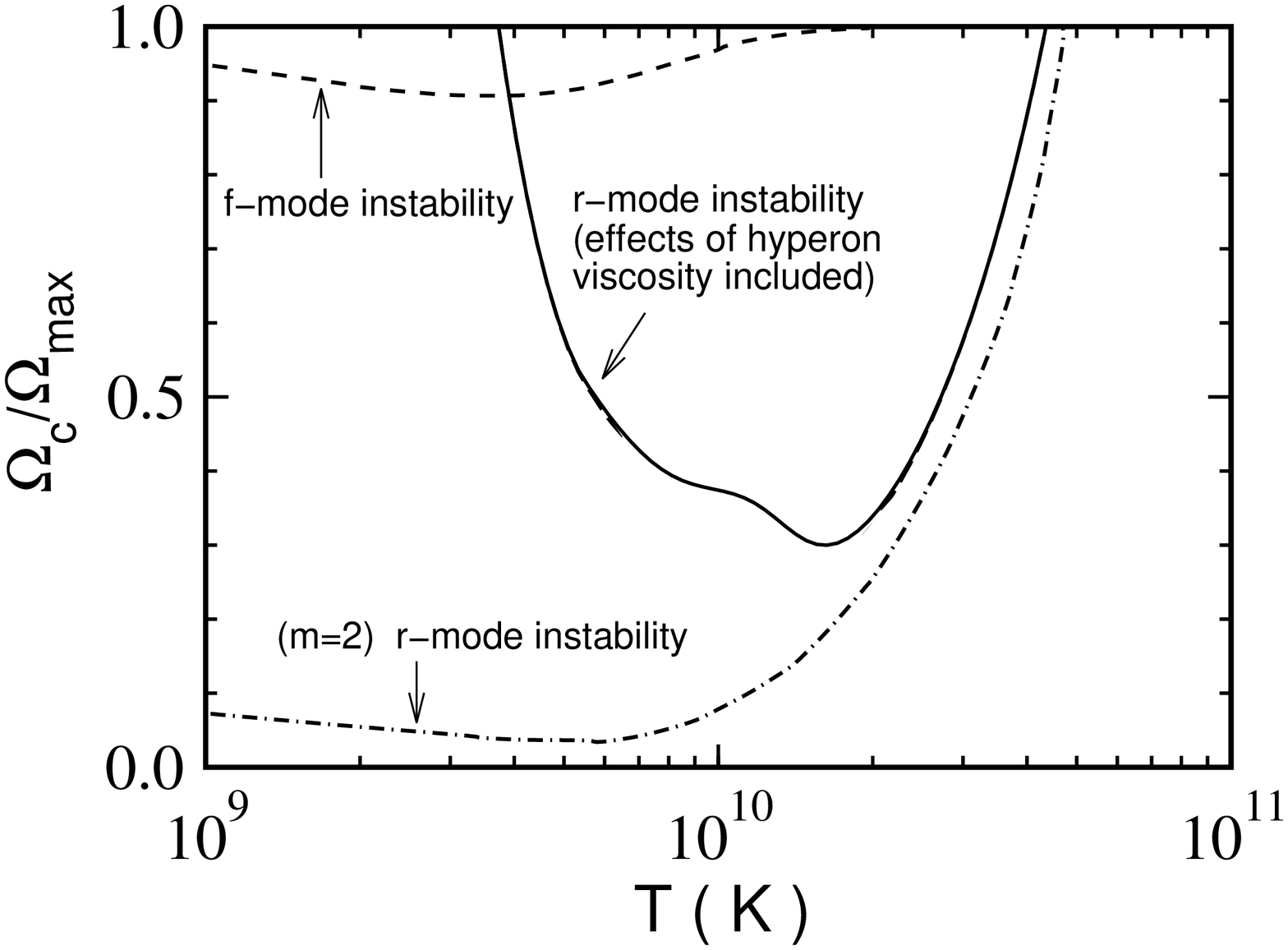} 
\caption{Temperature dependence of the critical angular velocity
$\Omega_{\rm c}$ of rotating neutron stars. The left panel shows the
gravitational radiation driven $f$-mode instability suppressed by
shear and bulk viscosity. Right panel: comparison of $f$-mode
instability with $r$-mode instability.  (Data from Refs.\
\cite{lindblom02:a,lindblom01:proc}.)}
\label{fig:OgrrvsT}
\end{center}
\end{figure}
transport of matter, does not damp the instability modes but rather
drives them. Viscosity plays the important role of damping these
instabilities at a sufficiently reduced rotational frequency such that
the viscous damping rate and power in gravity waves are comparable.
The most critical instability modes that are driven unstable by
gravitational radiation are $f$-modes and $r$-modes.  Figure
\ref{fig:OgrrvsT} shows the stable neutron star frequencies if only
$f$-modes were operative. One sees that hot as well as cold neutron
stars can rotate at frequencies close to mass shedding, because of the
large contributions of shear and bulk viscosity, respectively, for
this temperature regime.  The more recently discovered $r$-mode
instability may change the picture completely, as can be seen too from
Fig.\ \ref{fig:OgrrvsT}.  These modes are driven unstable by
gravitational radiation over a considerably wider range of angular
velocities than the $f$-modes (cf.\ dashed curve labeled ($m=2$)
$r$-mode instability). In stars with cores cooler than $\sim 10^9$~K,
on the other hand, the $r$-mode instability may be completely
suppressed by viscous phenomena so that stable rotation would be
limited by the $f$-mode instability again \cite{lindblom02:a}.

\begin{figure}[tb]
\begin{center} 
\parbox[t]{6.9cm}
{\includegraphics[width=6.5cm,angle=0]{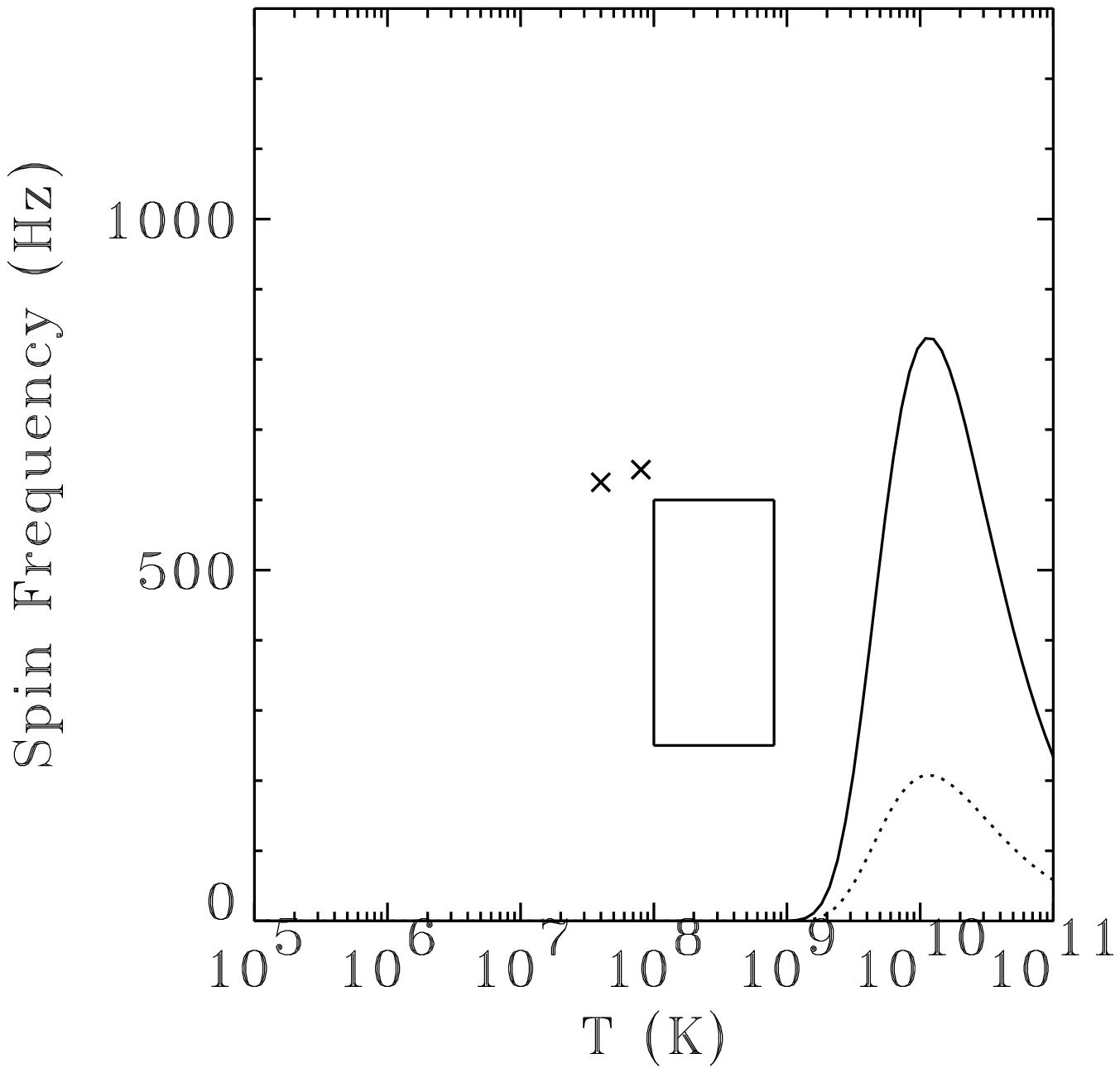}
\caption{Critical rotation frequencies versus stellar temperature for
CFL strange stars \cite{madsen00:b}.}
\label{fig:cfl}}
\ \hskip 1.0 cm \ 
\parbox[t]{6.9cm}
{\includegraphics[width=6.5cm,angle=0]{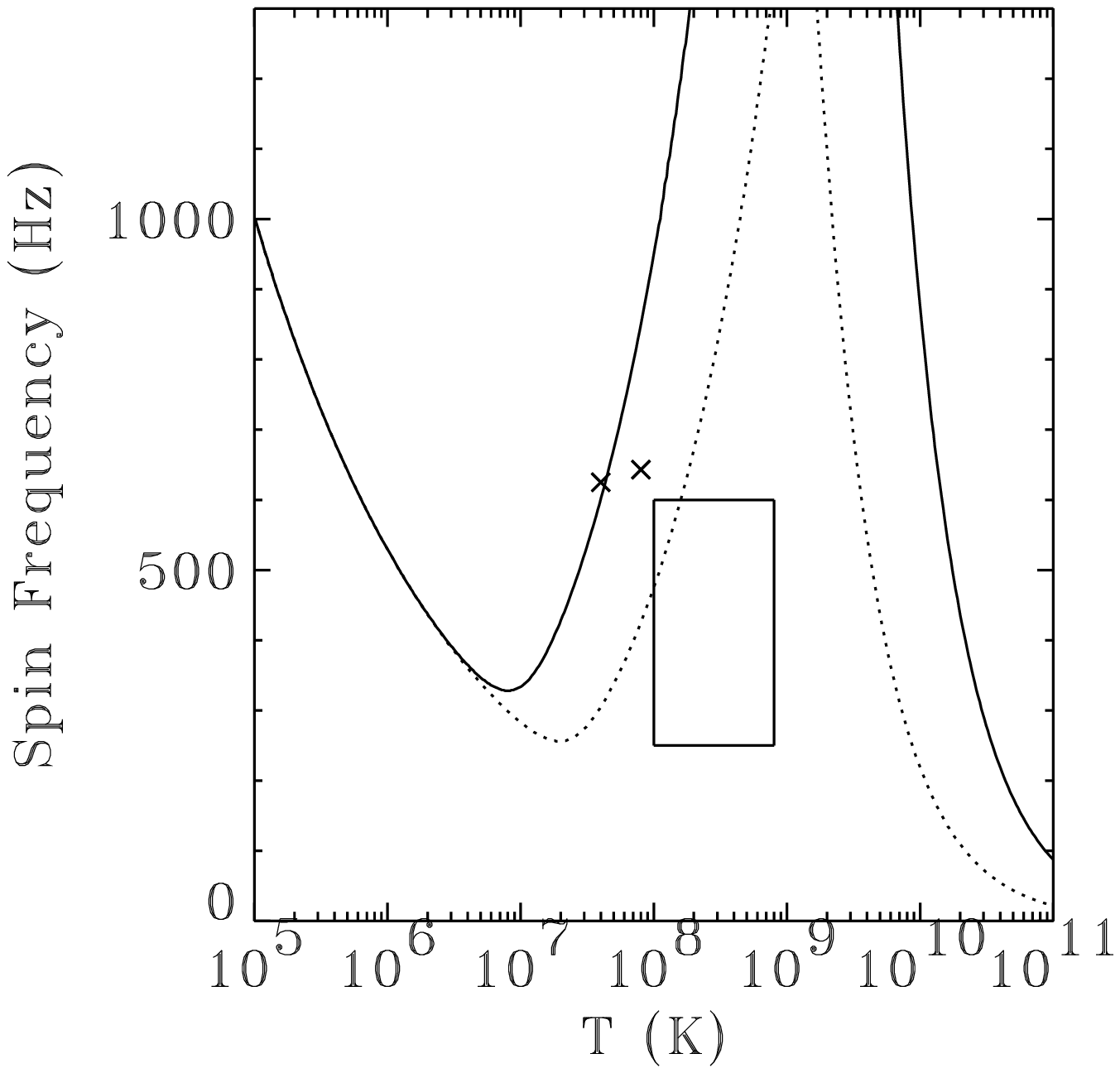} \caption{Same
as Fig.\ \protect \ref{fig:cfl}, but for 2SC quark stars
\cite{madsen00:b}.}
\label{fig:2sc}}
\end{center}
\end{figure}

Figures \ref{fig:cfl} and \ref{fig:2sc} are the counterparts to Fig.\
\ref{fig:OgrrvsT} but calculated for strange stars made of CFL and 2SC
quark matter, respectively \cite{madsen98:a,madsen00:b}. The $r$-mode
instability seems to rule out that pulsars are CFL strange stars, if
the characteristic time scale for viscous damping of $r$-modes are
exponentially increased by factors of $\sim \Delta/T$ as calculated in
\cite{madsen98:a}. An energy gap as small as $\Delta = 1$~MeV was
assumed. For much larger gaps of $\Delta \sim 100$~MeV, as expected
for color superconducting quark matter in the CFL phase, the entire
diagram would be $r$-mode unstable.  The full curve in Fig.\
\ref{fig:cfl} is calculated for a strange quark mass of $m_s =
200$~MeV, the dotted curve for $m_s=100$~MeV.  The box marks the
positions of most low mass X-ray binaries (LMXBs) \cite{klis00:a}, and
the crosses denote the most rapidly rotating millisecond pulsars
known. All strange stars above the curves would
spin down on a time scale of hours due to the $r$-mode instability, in
complete contradiction to the observation of millisecond pulsars and
LMXBs, which would rule out CFL quark matter in strange stars (see,
however, Ref.\ \cite{manuel04:a}). Figure \ref{fig:2sc} shows the
critical rotation frequencies of quark stars as a function of internal
stellar temperature for 2SC quark stars.  For such quark stars the
situation is less conclusive.  Rapid spin-down, driven by the $r$-mode
gravitational radiation instability, would happen for stars above the
curves.

\section{Astrophysical signals of quark deconfinement}\label{sec:aps}

\subsection{Isolated, rotating neutron stars}

Whether or not quark deconfinement occurs in neutron stars makes only
very little difference to their static properties, such as the range
of possible masses and radii, which renders the detection of quark
matter in such objects extremely complicated. This may be strikingly
different for isolated, rotating neutron stars which spin down, and
thus become more compressed, because of the emission of magnetic
dipole radiation and a wind of electron-positron pairs. For some
rotating neutron stars the mass and initial rotational frequency may
be just such that the central density rises from below to above the
critical density for dissolution of baryons into their quark
constituents. If accompanied by a pronounced shrinkage of the neutron
star, as is the case for the neutron star shown in the central panel
in Fig.\ \ref{fig:profiles}, the star's moment of inertia could change
dramatically. As shown in \cite{glen97:a}, the moment of inertia can
decrease so anomalously that it could even introduce an era stellar
spin-up that may last for $\sim 10^8$ years.  Since the dipole age of
millisecond pulsars is about $10^9$~years, one may roughly estimate
that about 10\% of the solitary millisecond pulsars could be in the
quark transition epoch and thus could be signaling the ongoing process
of quark deconfinement.  Changes in the moment of inertia reflect
themselves in the braking index, $n$, of a rotating neutron star, as
can be seen from \cite{weber05:a,glen97:a,spyrou02:a}
\begin{equation}  
  n(\Omega) \equiv \frac{\Omega\, \ddot{\Omega} }{\dot{\Omega}^2} = 3
- \frac{ I + 3 \, I' \, \Omega + I'' \, \Omega^2 } {I + I' \, \Omega}
\simeq 3 - \frac{ 3 \, I' \, \Omega + I'' \, \Omega^2 } {2\, I + I' \,
\Omega} \, ,
\label{eq:index}
\end{equation}  
where dots (primes) denote derivatives with respect to time
($\Omega$).  The last relation in (\ref{eq:index}) constitutes the
non-relativistic limit of the braking index \cite{glen95:a}. It is
obvious that these expressions reduce to the canonical limit $n=3$ if
$I$ is independent of frequency. Evidently, this is not the case for
\begin{table}[h]
\begin{center}
\caption{Dominant neutrino emitting processes in neutron star
cores if hyperons and quarks are absent \protect \cite{page05:a}.}
\label{tab:emis.core}
\begin{tabular}{|llcc|} 
\hline 
Name           &Process      &Emissivity               &  \\ 
               &             &(erg cm$^{-3}$ s$^{-1}$) &   \\
\hline 
\parbox[c]{3.4cm}{Modified Urca cycle\\ (neutron branch)} &
\rule[-0.4cm]{0.02cm}{0.85cm}
\bn n+n \rightarrow n+p+e^-+\bar\nu_e \\ n+p+e^- \rightarrow n+n+\nu_e \en  & 
$\sim 2\!\!\times\!\! 10^{21} \: R \: T_9^8$ & Slow \\
\parbox[c]{3.4cm}{Modified Urca cycle\\ (proton branch)}  &
\rule[-0.4cm]{0.02cm}{0.85cm}
\bn p+n \rightarrow p+p+e^-+\bar\nu_e \\ p+p+e^- \rightarrow p+n+\nu_e \en  & 
$\sim 10^{21} \: R \: T_9^8$ & Slow \\
Bremsstrahlung          &
\bn n+n \rightarrow n+n+\nu+\bar\nu \\ n+p \rightarrow n+p+\nu+\bar\nu \vspace{-0.2cm} \\
    p+p \rightarrow p+p+\nu+\bar\nu \en                                     & 
$\sim 10^{19} \: R \: T_9^8$  & Slow \\
\parbox[c]{3.5cm}{Cooper pair \\  formations}          &
\bn    n+n \rightarrow [nn] +\nu+\bar\nu \\  p+p \rightarrow [pp] +\nu+\bar\nu \en  & 
\bn \sim 5\!\!\times\!\! 10^{21} \: R \: T_9^7 \\ 
\sim 5\!\!\times\!\! 10^{19} \: R \: T_9^7 \en & Medium \\
Direct Urca cycle        & 
\rule[-0.4cm]{0.02cm}{0.85cm}
\bn n \rightarrow p+e^-+\bar\nu_e \\ p+e^- \rightarrow n+\nu_e \en              & 
$\sim 10^{27} \: R \: T_9^6$ & Fast \\
$\pi^-$ condensate &$n+<\pi^-> \rightarrow n+e^-+\bar\nu_e$  &
$\sim 10^{26} \: R \: T_9^6$  & Fast \\
$K^-$ condensate   &$n+<K^-> \rightarrow n+e^-+\bar\nu_e$  &

$\sim 10^{25} \: R \: T_9^6$  & Fast \\
\hline 
\end{tabular}
\end{center}
\end{table}
rapidly rotating neutron stars, and it also fails for stars that
undergo pronounced compositional changes (phase transitions) which
alter the moment of inertia significantly. Under favorable
\begin{table}[t]
\begin{center}
\caption{Dominant neutrino emitting processes in deconfined quark
matter \protect \cite{page05:a}.}
\label{tab:emis.quark}
\begin{tabular}{|llcc|} 
\hline 
     Name           &Process                 &Emissivity       &Efficiency    \\ 
                    &                        &(erg cm$^{-3}$ s$^{-1}$) &               \\
\hline
\parbox[c]{3.5cm}{Direct Urca cycle \\ ($ud$ branch)}         &
\rule[-0.4cm]{0.02cm}{0.85cm}
\bn u+e^- \rightarrow d+\nu_e \\ d \rightarrow u+e^-+\bar\nu_e \en
                                                                &     $\sim 10^{26} \: R \: T_9^6$    & Fast \\
\parbox[c]{3.5cm}{Direct Urca cycle \\ ($us$ branch)}         &
\rule[-0.4cm]{0.02cm}{0.85cm}
\bn u+e^- \rightarrow s+\nu_e \\ s \rightarrow u+e^-+\bar\nu_e \en
                                                                &     $\sim 10^{25} \: R \: T_9^6$    & Fast \\
\parbox[c]{3.5cm}{Modified Urca cycle \\ ($ud$ branch)}      &
\rule[-0.4cm]{0.02cm}{0.85cm}
\bn Q+u+e^- \rightarrow Q+d+\nu_e \\ Q+d \rightarrow Q+u+e^-+\bar\nu_e \en
                                                                &     $\sim 10^{21} \: R \: T_9^8$    & Slow \\
\parbox[c]{3.5cm}{Modified Urca cycle \\ ($us$ branch)}         &
\rule[-0.4cm]{0.02cm}{0.85cm}
\bn Q+u+e^- \rightarrow Q+s+\nu_e \\ Q+s \rightarrow Q+u+e^-\bar\nu_e \en
                                                                &     $\sim 10^{20} \: R \: T_9^8$    & Slow \\
Bremsstrahlungs &$Q_1+Q_2 \rightarrow Q_1+Q_2+\nu+\bar\nu$      &     $\sim 10^{19} \: R \: T_9^8$    & Slow \\
\parbox[c]{3.5cm}{Cooper pair \\ formations}          &
\bn    u+u \rightarrow [uu] +\nu+\bar\nu \\ d+d \rightarrow [dd] +\nu+\bar\nu \vspace{-0.2cm}
\\ s+s \rightarrow [ss] +\nu+\bar\nu \en  & 
\bn \sim 2.5\!\!\times\!\! 10^{20} \: R \: T_9^7 \\ \sim 1.5\!\!\times\!\! 10^{21} \: R \: T_9^7 \vspace{-0.2cm}
\\ \sim 1.5\!\!\times\!\! 10^{21} \: R \: T_9^7 \en & Medium \\
\hline 
\end{tabular}
\end{center}
\end{table}
circumstances, these changes in $I$, originating from the transition
of confined hadronic matter into quark matter, may cause the braking
index to deviate dramatically from 3 in the vicinity of the star's
frequency where the phase transition to quark matter occurs.  The
changes in $I$ may even be so pronounced that $n(\Omega) \rightarrow
\pm \infty$ at the transition point
\cite{glen97:book,weber99:book,weber05:a,glen95:a}. Such dramatic
anomalies in $n(\Omega)$ are not known for conventional neutron stars
(see left panel in Fig.\ \ref{fig:profiles}), because their radii and
thus moments of inertia appear to vary smoothly with $\Omega$
\cite{weber99:book,weber05:a}. A counterexample to this, however, is
discussed in \cite{zdunik04:a}. The future astrophysical observation of
strong anomalies in the braking behavior of isolated pulsars could
thus be cautiously interpreted as a possible astrophysical signal for
quark deconfinement in neutron stars.

\subsection{Accreting neutron stars}
\label{sec:ano2}

Accreting x-ray neutron stars provide a very interesting contrast to
the spin-down of isolated neutron stars. These x-ray neutron stars are
being spun up by the accretion of matter from a lower-mass ($M \ls 0.4
\msun$), less-dense companion.  If the critical deconfinement density
\begin{figure}[htb]
\begin{center}
\includegraphics*[width=0.6\textwidth,angle=0,clip]{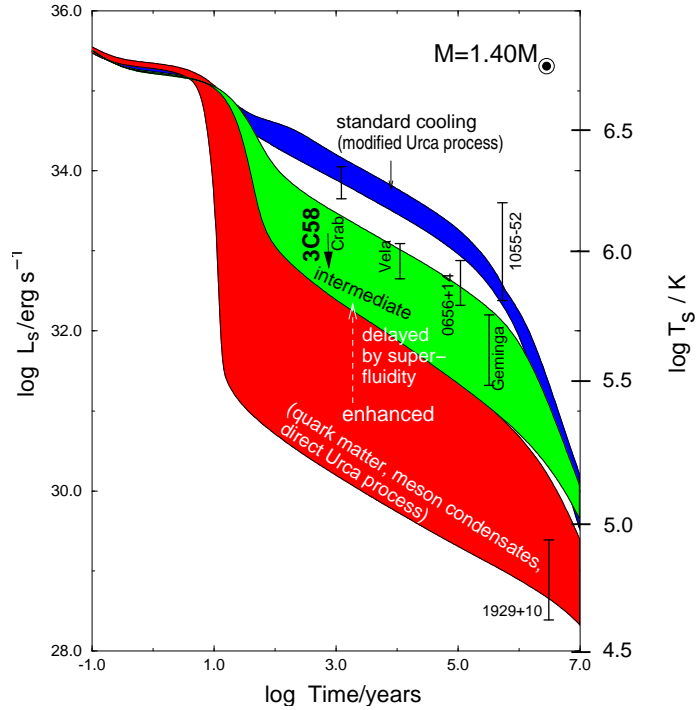}
\caption{Cooling behavior of a $1.4\,\msun$ neutron star for competing
assumptions about the properties of superdense matter. Three distinct
cooling scenarios, referred to as standard, intermediate, and enhanced
can be distinguished. The band-like structures reflect the
uncertainties inherent in the stellar EoS \protect
\cite{weber99:book,weber05:a}.}
\label{fig:cool} 
\end{center}
\end{figure}
falls within that of the canonical pulsars, quark matter could already
exist in them but will be spun out of such stars as their frequency
increases during accretion.  This scenario has been modeled in
\cite{glen01:a}, where it was found that quark matter remains
relatively dormant in the core of a neutron star until the star has
been spun up to frequencies at which the central density is about to
drop below the threshold density at which quark matter exists. As
known from the discussion above, this could manifest itself in a
significant increase of the star's moment of inertia. The angular
momentum added to a neutron star during this phase of evolution is
therefore consumed by the star's expansion, inhibiting a further
spin-up until the star's quark matter content has been completely
converted into a mixed phase of hadrons and quarks.  Such accreters,
therefore, tend to spend a greater length of time in the critical
frequencies than otherwise. For canonical accretion rates of
$10^{-10}\, \msun$/year the time span can be on the order of
$10^9$~years. Hence, from this scenario, one would expect a greater
number of accreting x-ray neutron stars that appear near the same
frequency. Evidence that accreting neutron stars pile up at certain
frequencies, which are well below the mass shedding limit, is provided
by the spin distribution of accreting millisecond pulsars in 57 Tuc
and neutron stars in low mass X-ray binaries observed with the Rossi
X-ray Timing Explorer (RXTE). The proposed limiting mechanisms
responsible for this behavior is generally attributed to gravity-wave
emission caused by the $r$-mode instability, or by a small stellar mass
quadrupole moment
\cite{bildsten98:a,andersson00:a,chakrabarty03:a}. Supplemental to
these explanations, quark reconfinement (or, more generally, strong
first-order like phase transition) may be linked to this phenomenon as
well \cite{glen00:b,glen01:a,chubarian00:a,poghosyan01:a}.

\section{Cooling of neutron stars}\label{sec:cooling}

The predominant cooling mechanism of hot (temperatures of several
$\sim 10^{10}$~K) newly formed neutron stars immediately after
formation is neutrino emission, with an initial cooling time scale of
seconds. Already a few minutes after birth, the internal neutron star
temperature drops to $\sim 10^9$~K. Photon emission overtakes neutrino
emission when the internal temperature has fallen to $\sim 10^8$~K,
with a corresponding surface temperature roughly two orders of
magnitude smaller. Neutrino cooling dominates for at least the first
$10^3$ years, and typically for much longer in standard cooling
(modified Urca) calculations. The dominant neutrino emitting processes
in neutron star matter are summarized in Tables \ref{tab:emis.core}
and \ref{tab:emis.quark}.  Figure \ref{fig:cool} shows the outcome of
cooling calculations performed for a broad collection of equations of
state \cite{weber99:book,weber05:a} and competing assumptions about
the dominant neutrino emitting processes. For recent overviews of
neutron star cooling, see, for instance, Refs.\
\cite{page05:a,yakovlev04:a}. We also refer to D.\ Blaschke's
contribution contained elsewhere in this volume. 

\section{Summary}

It is often stressed that there has never been a more exciting time in
the overlapping areas of nuclear physics, particle physics and
relativistic astrophysics than today.  This comes at a time where new
orbiting observatories such as the Hubble Space Telescope (HST), Rossi
X-ray Timing Explorer, Chandra X-ray satellite, and the X-ray Multi
Mirror Mission (XMM) have extended our vision tremendously, allowing
us to observe compact star phenomena with an unprecedented clarity
and angular resolution that previously were only imagined.  On the
Earth, radio telescopes (Arecibo, Green Bank, Parkes, VLA) and
instruments using adaptive optics and other revolutionary techniques
have exceeded previous expectations of what can be accomplished from
the ground. Finally, the gravitational wave detectors LIGO, LISA, and
VIRGO are opening up a window for the detection of gravitational waves
emitted from compact stellar objects such as neutron stars and black
holes. This unprecedented situation is providing us with key
information on compact stars, which are the only physical objects in
which cold and dense baryonic matter is realized in nature.  As
discussed in this paper, a key role in compact star physics is played
by strangeness. It alters the masses, radii, cooling behavior, and
surface composition of neutron stars.  Other important observables may
be the spin evolution of isolated neutron stars and neutron stars in
low-mass x-ray binaries. All told, these observables are key in
exploring the phase diagram of dense nuclear matter at high baryon
number density but low temperature, which is not accessible to
relativistic heavy ion collision experiments.

\end{document}